\documentclass{emulateapj}
\usepackage{graphicx}

\begin{document}

\slugcomment{AJ, 2008 March, in press}

\title{Mass Outflow and Chromospheric Activity of Red Giant Stars in Globular Clusters I: M15}

\author{Sz. Meszaros\altaffilmark{1,2,3}, A. K. Dupree\altaffilmark{1,4}, and A. Szentgyorgyi\altaffilmark{1,5}}

\altaffiltext{1}{Harvard-Smithsonian Center for Astrophysics, Cambridge, MA 02138}
\altaffiltext{2}{Department of Optics and Quantum Electronics, University of Szeged, 6701 Szeged, Hungary}
\altaffiltext{3}{e-mail address: meszi@cfa.harvard.edu}
\altaffiltext{4}{e-mail address: adupree@cfa.harvard.edu}
\altaffiltext{5}{e-mail address: saint@cfa.harvard.edu}

\begin{abstract}
High resolution spectra of 110 selected red giant stars in the globular cluster M15 (NGC~7078) were obtained with 
Hectochelle at the MMT telescope in 2005 May, 2006 May, and 2006 October. Echelle orders containing 
H$\alpha$ and \ion{Ca}{2}~H~$\&$~K are used to identify emission and line asymmetries characterizing motions in the extended
atmospheres. Emission in H$\alpha$ is detected to a luminosity of $log (L/L_{\odot})=2.36$, in this very metal deficient
cluster, comparable to other studies, suggesting that appearance of emission wings is independent of stellar metallicity. 
The faintest stars showing H$\alpha$ emission appear to lie on the asymptotic giant branch (AGB) in M15. 
A line-bisector technique for H$\alpha$ reveals outflowing velocities in all
stars brighter than $log (L/L_{\odot})=2.5$, and this outflow velocity increases with stellar luminosity, indicating the mass
outflow increases smoothly with luminosity. Many stars lying low on the AGB show exceptionally high outflow velocities 
(up to 10$-$15 km s$^{-1}$) and more velocity variability (up to 6$-$8~km~s$^{-1}$), than red giant branch (RGB) 
stars of similar apparent magnitude. High velocities in M15 may be related to the low cluster metallicity. 
Dusty stars identified from \textit{Spitzer} 
Space Telescope infrared photometry as AGB stars are confirmed as cluster members by radial velocity measurements,
yet their H$\alpha$ profiles are similar to those of RGB stars without dust. If substantial mass loss creates the 
circumstellar shell responsible for infrared emission, such mass loss must be episodic.
\end{abstract}


\section{Introduction}

The well-known second parameter problem in globular clusters \citep{sandage02}, 
in which a parameter other than metallicity, affects the morphology of the horizontal branch, remains unresolved. 
Metallicity, as first noted by \citet{sandage01}, remains the principal parameter, but pairs of clusters,
with the same metallicity, display quite different horizontal branch morphologies thus challenging the canonical models of 
stellar evolution and leading to the need for a `Second Parameter'.
Cluster ages have been  examined in many studies \citep{searle01, lee02, stetson01, lee01, sarajedini01, sarajedini02} 
and in addition, many other suggestions for the `second parameter(s)' have been proposed,
including: total cluster mass; stellar environment (and possibly free-floating
planets); primordial He abundance; post-mixing surface
helium abundance; CNO abundance; stellar rotation; and mass loss \citep{catelan01, catelan02, sills01, soker01, sweigart01, buonanno01, peterson03, 
buonanno02, recio01}. Many authors \citep{vandenberg01, lee02, catelan01} have proposed that more than one 
second parameter may exist in addition to age. One parameter may be mass loss which, as \citet{catelan01} 
notes, remains an `untested second-parameter candidate'.

This spectroscopic study addresses the presence of mass loss from luminous stars in the globular
cluster M15. Subsequent papers will include other clusters. Although
stellar evolution theory predicts that low-mass Population II stars
ascending the red giant branch (RGB) for the first time must lose
mass \citep{renzini01, sweigart02}, few observations have
identified the ongoing mass loss process. Evidence
from the period-luminosity relation for RR Lyrae stars suggests that the
luminosity variations can be accommodated theoretically if mass loss $\sim 0.2-0.4 \ M_{\odot}$
has occurred \citep{fusi01, christy01}. Circumstellar CO 
emission in M-type irregular and semiregular asymptotic giant branch (AGB)-variables implies mass loss rates on the AGB 
$\sim 10^{-7} - 10^{-8} \ M_{\odot} \ yr^{-1}$ \citep{olofsson01}. Indirect
evidence of mass loss processes would be detection of an intracluster medium. These efforts have been marginally
successful. Diffuse gas ($< \ 1 \ M_{\odot}$) was 
suggested in NGC~2808 through the detection of 21$-$cm H line emission \citep{faulkner01}, but has remained unconfirmed.
Ionized intracluster gas was found
in the globular cluster 47 Tucanae by measuring the radio dispersion of millisecond pulsars in the cluster \citep{freire01}. 
The central electron density was derived ($n_e=0.067 \ \pm 0.0015$~cm$^{-3}$) and found to be 
two orders of magnitude higher than the ISM in the vicinity of 47 Tuc \citep{taylor01}. \citet{freire01} determined the electron
density in M15 using four millisecond pulsars to be higher ($n_e\sim0.2$~cm$^{-3}$) than in 47 Tuc.

Indirect evidence of mass loss processes comes also from infrared observations.
\citet{origlia01} using ISOCAM images found a mid-IR excess associated with
giants in several globular clusters and attributed to dusty circumstellar
envelopes. The first detection of intracluster dust in M15 was made by \citet{evans01} from the
analysis of far infrared imaging data obtained with the ISO instrument ISOPHOT. \citet{loon01} also 
presented a tentative detection of 0.3 $M_{\odot}$ of neutral hydrogen in M15. \citet{smith02} placed 
an upper limit of 0.4 $M_{\odot}$ for the molecular gas in M15 from CO observations
with the 15-m James Clerk Maxwell Telescope on Mauna Kea.
Using the \textit{Spitzer} Space Telescope, \citet{boyer01} detected a population of dusty red giants 
near the center of M15. Observations from \textit{Spitzer} with the Multiband Imaging Photometer for \textit{Spitzer} (MIPS) 
also revealed the intracluster medium (ICM) discovered by 
\citet{evans01} near the core of the globular cluster. As \citet{origlia01} noted, the infrared detections may only
be tracing the outflowing gas and may not be related to the driving mechanisms for the wind.
More recently \citet{origlia02} identified dusty RGB stars in 47 Tuc and derived an empirical mass loss law for Population II
stars. Mass loss rates derived from these observations showed that the mass loss increases with luminosity and possibly 
episodic.

High resolution stellar spectroscopy allows the direct detection of mass outflow from the red giants
themselves. Emission in the wings of H$\alpha$ lines in the
spectra of globular cluster red giants was first described in detail by \citet{cohen01}. 
Later observations revealed that emission in H$\alpha$ is common in globular clusters and night-to-night variations
can occur \citep{mallia01, peterson01, peterson02, cacciari02, gratton01}. These
studies have shown that most of the stars brighter than $log (L/L_{\odot}) \approx 2.7$ exhibit H$\alpha$ emission wings.
The emission itself is likely not a direct indicator of mass loss, because 
emission can arise from an optically thick stellar chromosphere surrounding the star \citep{dupree01}. 
Variation of the strength of emission can also be affected by stellar pulsation \citep{smith01}. Better mass 
flow indicators in the optical are
the line coreshifts or asymmetries of the H$\alpha$ or \ion{Ca}{2}~H$\&$K profiles and emission features. 
Red giants in globular clusters (M22 and Omega Centauri) were found to have velocity
shifts less than 14~km~s$^{-1}$ in the cores of H$\alpha$ relative to the photospheric lines \citep{bates01, bates02}. 
These results were similar to metal-poor field giants, where
only giants brighter than $M_V=-1.7$ have emission wings and the line shifts were $< \ 9$~km~s$^{-1}$ 
\citep{smith01}
indicating very slow outflows and inflows in the chromosphere. For globular clusters, \citet{lyons01} discussed the
H$\alpha$ and Na~I~D line profiles for a sample of 63 RGB stars in M4, M13, M22, M55, and $\omega$~Cen. 
The coreshifts
were less than 10~km~s$^{-1}$, much smaller than the escape velocity from the stellar atmosphere at $2 \ R_{*}$ 
($\approx 50-70$~km~s$^{-1}$). \citet{dupree02} studied 2 RGB stars in NGC~6752 and found that the \ion{Ca}{2}~K and 
H$\alpha$ coreshifts were
also low (less than 10~km~s$^{-1}$). However, asymmetries in the Mg~II lines showed a stellar wind with a
velocity of $\approx 150$~km~s$^{-1}$, indicative of a strong outflow. 
Mg~II lines are formed higher
in the atmosphere than H$\alpha$ and \ion{Ca}{2}~K, which suggests that the stellar wind becomes detectable near the top of 
the chromosphere. 
A detailed study was carried out by \citet{cacciari01}, who observed 137 red giant stars in NGC~2808.
Most of the stars brighter than $log (L/L_{\odot})=2.5$ clearly showed 
emission wings in H$\alpha$. The velocity shift of the H$\alpha$ line core compared to the photosphere is less than 
$\approx 9$~km~s$^{-1}$. Outward motions were also found in both Na~I~D and \ion{Ca}{2}~K profiles. 

This paper discusses high-resolution spectroscopy of the H$\alpha$ and \ion{Ca}{2}~H$\&$K lines of
red giant stars in M15. Our deep sample of M15 giants in this metal poor cluster ([Fe/H]=$-$2.26) offers a good
comparison to other studies of more metal rich clusters. Also, the high radial
velocity of M15 ($-$105~km~s$^{-1}$) minimizes contamination by interstellar absorption in the profiles of resonance lines.

\section{Observations and Data Reduction}

Observations of a total of 110 red giant stars in M15 were made in 2005 May, 2006 May, and 2006 October with Hectochelle on
the MMT \citep{szentgyorgyi01}. Hectochelle uses 240 fibers each of which subtends $\sim$ 1.5 arcsec in the sky. Hectochelle
covers 1 degree of sky, yet the apparent diameter of M15 is only $\sim$10 arc minutes, so that about 50$-$60 red giants
could be measured in M15 for each configuration. The requirement that fibers can not be
placed closer than 2 arcsec apart further constrains the target selection. In addition, we wanted to search for variability 
which led to multiple visits for many targets over the 17 month span. Software ({\it xfitfibs}) has been developed 
at CfA to optimize the fiber configuration with specified priorities and requirements.

Targets were chosen from the catalog of \citet{cudworth01} to have a high probability ($>95\%$) of membership and
to provide smooth coverage of the RGB and AGB within the constraint of the fiber placement on the sky. 
The color magnitude diagram (CMD) of the observed cluster members can be seen in Figure 1 and they are listed in Table 1. 
Coordinates of the stars were taken from the 2MASS catalog \citep{skrutskie01} and used to position the fibers.
Additional targets from Cudworth's list with lower membership probability and field targets from the 2MASS catalog 
were included. Many fibers
were placed on the blank regions of the sky in order to measure the sky background in detail. 
The sky fibers were equally distributed in the observed field to cover a large area around the cluster.
Since Hectochelle is a single-order instrument, two orders were selected with order-separating filters: 
H$\alpha$ (region used for analysis $\lambda \lambda \ 6475-6630$)\footnote{The 2005 May observation had less wavelength 
coverage. See following text.}, and \ion{Ca}{2}~H$\&$K (region used for analysis $\lambda \lambda \ 3910-3990$) to give 
155\AA \ centered on the principal spectral features in H$\alpha$ and 80\AA \ in \ion{Ca}{2}~H$\&$K.
The spectral resolution was about 34,000 as measured by the FWHM of the ThAr emission lines in the comparison lamp. 
Exposures in each of the two orders are summarized in Table 2.

Data reduction was done using standard IRAF spectroscopic packages. The IRAF package {\it ccdproc} performed 
the trimming and 
the overscan correction and made the bad pixel mask. The average bias image was subtracted 
from every spectrum. Correcting with the dark images was not necessary because even in the 40 minute dark 
exposures the intensity was very low [3~$-$~4 analog digital units (ADU) per pixel]. 
To find and trace the apertures, ten flat images were taken with the continuum 
lamp of 10 seconds exposure time each. The focal plane of Hectochelle consists of a mosaic of 2 CCDs that are slightly
misaligned. The aperture finding algorithm fails near the crack between the two CCDs, so manually editing the 
apertures and reordering them was 
necessary. Some apertures were deleted from the edges of the CCD and a total of 240 orders was extracted. 
To correct for the pixel-to-pixel variations, the averaged continuum flat exposures for each configuration were fitted 
with a 21st order spline function (using the IRAF task {\it apflatten}) and used to divide the corresponding object 
spectra by the normalized flat. A region 13$-$pixels wide was used for the aperture extraction. 
Wavelength standards, using Thorium-Argon (ThAr) hollow cathode lamps were taken to define the wavelength scale; 
each ThAr image had 900 seconds of exposure time and was taken at the beginning of the night. 
We identified 15~$-$~20 strong ThAr emission lines in the first aperture, then propagated these identifications to 
every other aperture manually to check the accuracy of the fit.
During the calibration the rms of the wavelength fit had to be
between 0.01~$-$~0.002\AA \ to reach the theoretical resolution of the spectrograph. If the error of the fit is
larger than this, it will be comparable to the expected width of the ThAr features (0.1~$-$~0.2\AA) and increase the error 
of the wavelength solution.
The continuum flat images were used to correct the throughput for each aperture using a region close to
the CCD center containing 5~$-$~7 neighboring fibers. An average of the selected continuum flat apertures was taken and 
divided into all other apertures in the same exposure and was used to correct the vignetting and fiber-to-fiber 
throughput deviation. 

The extracted spectra also contain sky background which had to be subtracted. Some of the sky apertures
showed weak H$\alpha$ and other photospheric lines suggesting very faint stars in those positions. Also very bright stars
can cause scattered light in the neighboring apertures on the CCD, but this becomes visible only if the aperture has more 
than 8000~$-$~10000 ADUs per pixel. The brightest stars reached 10000 ADUs per pixel in the 40 minute exposures and 
some apertures contained very low level scattered light.
Every sky aperture was checked carefully and those where faint stars or scattered light were found were discarded.
A median filtered sky was used for the subtraction, but sky subtracted skies frequently contained additional counts, 
which changed aperture by
aperture and by wavelength. In Figure 2, sky intensity versus wavelength and aperture is plotted. 
Between aperture numbers 100 and 150, especially at longer wavelengths, the apertures have higher intensity. 
The dark images did not show high intensity features and the intensity pattern in Figure 2 is currently not understood. 
To subtract the sky, the images were divided into 3 different aperture sections, and the sky subtraction was done with one 
of two methods. In the first and
third segments (corresponding to aperture numbers 0~$-$~100 and 150~$-$~240), the intensity appears constant as a function of 
aperture number and wavelength so the median filtered spectrum could be used for subtraction. 
The middle region spanned aperture numbers 101 to 149. 
In this region the sky was subtracted from every target aperture using the average of 3 closest sky apertures on the 
CCD itself. 

In the 2005 May spectra of H$\alpha$, the filter's central wavelength was offset by $\sim$ 80\AA \ 
placing the H$\alpha$ line near the long wavelength end of the CCD. Fluxes at wavelengths shorter than H$\alpha$
were abnormally low, because the grating was so far off the blaze angle. The wavelength regions spanned by the OB25
filter differed between the 2005 and 2006 observations; however both contained the H$\alpha$ line and photospheric
lines.

\section{Stars with H$\alpha$ Emission, Radial Velocities}

\subsection{The Color Magnitude Diagram and Physical Parameters}

We observed a total of 110 different red giant stars in M15 and found 29 with H$\alpha$ emission. About half of them were 
observed more than once. Emission above the continuum in the H$\alpha$ profile can be seen in Figures 3~$-$~6. 
For comparison each figure includes a star that exhibits no emission. 
The color-magnitude diagram (CMD) for each night of observation appears in Figure 7. 
On 2005 May 22, emission is found in stars of V=14.48 
and brighter, corresponding to $M_V=-0.89$, using the apparent distance modulus $(m-M)_V=15.37$ from \citet{harris01}.
Stars in M15 that show emission occurred at different magnitude limits on different dates of observation 
($M_V=-1.17$ on 2006 May 11, $M_V=-0.99$ on 2006 October 4 and $M_V=-1.68$ on 2006 October 7). 
Studies of metal deficient field giants found emission in objects
brighter than $M_V=-1.7$ \citep{smith01}, whereas in the metal rich cluster NGC~2808 the detection threshold for emission
was set at $M_V=-1.0$ \citep{cacciari01}. It is well documented that the presence of H$\alpha$
varies with time, and this appears to be the most likely explanation of the differences in the detection level of the 
H$\alpha$ emission.

For comparison on a luminosity scale, unreddened colors for M15 stars were calculated taking $E(B-V)=0.10$, 
$E(V-K)=2.75E(B-V)$, and apparent
distance modulus $(m-M)_V=15.37$ \citep{cardelli01, harris01}. The effective temperatures, bolometric 
corrections, and luminosities were obtained from the $V-K$ visual colors (Table 3) using the empirical calibrations by 
\citet{alonso01,alonso02} and the cluster average metallicity [Fe/H]=$-2.26$ \citep{harris01}. 

Stars brighter than 
$log (L/L_{\odot})=2.36$ can exhibit emission, and all together $\sim 46\%$ of these show H$\alpha$ emission. The asymmetry 
of the H$\alpha$ emission wings was noted for emission above the continuum level where B represents the strength of the short
wavelength (blue) emission and R denotes the strength of the long wavelength (red) emission.
Figure 7 shows the CMD for our targets where the asymmetry of the H$\alpha$ line is indicated.
The frequency of H$\alpha$ emission increases with the stellar luminosity, however the line asymmetry is not correlated with
color and luminosity. Stars with B$<$R and B$>$R seem equally distributed in luminosity. 
Among stars with emission, the majority ($\sim 75\%$) exhibit emission wings with B$>$R, a signature generally considered 
to indicate inflow of material. 

Spectra of 29 stars with H$\alpha$ emission were obtained in both 2005 and 2006. 
All but two of these stars showed significant changes in the line
emission which either appeared, or vanished, or changed asymmetry (see Table 4, and Figures 3$-$6).

\subsection{Radial Velocities}

To measure accurate radial velocities we chose the cross-correlation method using the IRAF task {\it xcsao}. 
Using the ATLAS \citep{kurucz01} code, we synthesized the
spectrum of a red giant star, K341, in M15. This is a bright star with a high quality spectrum, thus the comparison between 
the observed and modeled spectrum is optimum. The physical parameters
of our template spectrum 
were the following: $T_{eff}=4275$~K, $log \ g=0.45$, $v_{turb}=0$ km~s$^{-1}$, $v_{macro}=0$ km~s$^{-1}$, 
[Fe/H]=$-$2.45, [Na/Fe]=0.01, [Si/Fe]=0.4, $\ $ [Ca/Fe]=0.56, [Ti/Fe]=0.57, [Ba/Fe]=0.2 \citep{sneden01}. This spectrum 
is computed in LTE and a chromosphere was not included in the atmospheric model.
The comparison between the template in our cross-correlation and the observed Hectochelle spectrum can be seen in Figure 8.
The region selected for the cross-correlation spanned 6480\AA \  to 6545\AA \ purposely omitting the H$\alpha$ line. 
The telluric and photospheric lines were identified using the synthesized spectrum of K341. 
In this region there are many telluric lines of water vapor. 
These lines appear in the cross-correlation function 
profile as an additional peak, well separated from the cluster velocity, and so
the measured stellar radial velocity is not affected. To verify our radial velocities, we cross-correlated a narrow region 
$\lambda \lambda \ 6480-6500$ where no strong telluric lines are found.
Cross-correlating this narrow region results in the same radial velocity as from the broader window, 
but with a larger (1$-$2~km~s$^{-1}$) error.

The H$\alpha$ spectra of our targets were also cross-correlated against several hundred spectra 
calculated by \citet{coelho01} covering temperatures between 3500 and 7000 K and
metallicities between [Fe/H]=$-$2.5 and +0.5. These velocities from the Coelho spectra agreed within 1 to 2~km~s$^{-1}$ 
with our earlier determination using only the K341 template, because the same photospheric Fe and
Ti absorption lines can be found in all the spectra.

There is good agreement between our radial velocities from the 2005 data and those of \citet{gebhardt01} and \citet{peterson04} (see
Figure 9, top panels). \citet{gebhardt01} used an Imaging Fabry-Perot Spectrophotometer with the Sub-arcsecond Imaging 
Spectrograph
on the Canada-France-Hawaii Telescope and observed 1534 stars in M15 with velocity errors between 0.5 and 10~km~s$^{-1}$. 
\citet{peterson04} used 
echelle spectrographs on the MMT, the 1.5 m Tillinghast reflector of the Whipple Observatory on Mount Hopkins, and the 
4$-$m telescope of Kitt Peak National Observatory. 
\citet{peterson04} quote an average error of 1~km~s$^{-1}$, but the stars in common with our sample 
have larger errors (1$-$2~km~s$^{-1}$). 
However the Hectochelle velocities from 2006 display a systematic offset from the 2005 measurements of
the same stars (see Figure 9, lower panels). This offset amounts to $+1.9 \ \pm 0.5$~km~s$^{-1}$ and 
$+0.9 \ \pm 0.5$~km~s$^{-1}$ for 2006 May and Oct 2006 respectively, and the data in Table 5 were corrected for this 
systematic offset.  
The radial velocities of the sky emission lines show the same effect. In 2005, all of the sky emission
lines were at 0~km~s$^{-1}$ and in 2006 May were at $-$2~km~s$^{-1}$, yet the wavelength calibration
of the 2006 data appears to be as accurate as 2005. The source of this offset comes from revisions made to the calibration 
system of Hectochelle that changed the illumination in the spectrograph. 
The amount of the offset is small, and does not affect determination of cluster membership.
The average cluster radial velocity was calculated using velocity-corrected data from all four observations. Our
value is $-105.0 \ \pm 0.5$~km~s$^{-1}$, which is slightly lower than the cluster radial velocity 
($-107.0 \ \pm 0.2$~km~s$^{-1}$) quoted in the \citet{harris01} catalog. 

Previous studies suggested that several of our sample stars are binaries. 
Significant velocity variations, $\approx$~6.5~km~s$^{-1}$, for K47 were found by \citet{soderberg01}, but our measurements 
showed only 0.9~km~s$^{-1}$ variation between 2005 May and 2006 May. This change lies within the measurement errors 
($\approx$~1~km~s$^{-1}$). K757 and K825 were suggested as binaries by \citet{sneden02} from the asymmetric line profiles;
weak satellite wings were visible for nearly all spectral lines. 
We have only one observation of K825, but K757 changed by 6.2~km~s$^{-1}$, which could indicate that this star is a binary. 
\citet{drukier01} found 17 cluster members of M15 showing possible radial velocity variability. Four of these stars were 
observed with Hectochelle, but only one of them, K92, was observed more than once. This star showed 1.4~km~s$^{-1}$
variability between 2005 May 22 and 2006 October 7, but the error of these observations was close to 1~km~s$^{-1}$.
Six additional stars showed velocity changes larger than 2~km~s$^{-1}$, which could indicate these stars are binaries: 
B5 (6.8~km~s$^{-1}$), B30 (2.9~km~s$^{-1}$), K757 (6.2~km~s$^{-1}$), K1084 (2.6~km~s$^{-1}$), K1097 (2.1~km~s$^{-1}$), 
and K1136 (3.0~km~s$^{-1}$).

Some of the M15 targets selected from the proper motion study of M15 \citep{cudworth01} turned out to have substantially different
radial velocities from the cluster average (see Table 6) and are not likely to be members of the cluster. 
These stars are not included in the spectroscopic analysis.

\subsection{Bisector of H$\alpha$ Lines}

The core of the H$\alpha$ line is formed higher in the stellar chromosphere than the line wings and is expected to give an
indication of the atmosphere dynamics.
We first measured the position of the H$\alpha$
absorption line core relative to photospheric lines using the IRAF task {\it splot}. 
We found an error in the wavelength scales depending on the aperture number that prevented measurement of the core
offset of H$\alpha$ to better than $\pm$~2~km~s$^{-1}$. At present, we believe that the different light path of the ThAr
comparison lamp from that of the stellar spectra causes this error, which appeared as a variable `stretching' of the
wavelength scale dependent on aperture and zenith distance of M15\footnote{Since this discovery, new observational procedures
have been instituted with Hectochelle using sky spectra to eliminate these effects.}. 

Thus, a better approach to the velocity differences
consists of measuring the line asymmetry using a line bisector. The difference between the centers of the line core and of the
line near the continuum level gives a measure of the atmospheric dynamics through the chromosphere.
To accomplish this, the line profile was divided into 20 sectors in normalized flux. The top sector was usually between 0.7 
and 1.0 of the continuum in the normalized spectrum, the lowest sector was placed 0.01~$-$~0.05 above the lowest value of 
the line depending on its signal-to-noise ratio. 
The velocities of the H$\alpha$ bisector asymmetry ($v_{bis}$) are calculated in the following way: the top and the
bottom 3 sectors are selected, the wavelength average of each sector is calculated, then subtracted one from another and
changed to a velocity scale. The bisector velocities, $v_{bis}$, are shown in Figure 10 and listed in Table 7. 
A negative value corresponds to an outflowing velocity.
Probable errors for these measurements were formally calculated and range from 0.5~km~s$^{-1}$ to 1.5~km~s$^{-1}$ for the 
brightest to faintest stars respectively. Stars fainter than $log (L/L_{\odot})=2.5$ (V~$\approx$~14.20) 
did not show H$\alpha$ asymmetry and $v_{bis}$ was nearly zero. 
Stars brighter than $log (L/L_{\odot})=2.5$ (V~$\approx$~14.20) showed asymmetry and in almost every case the line profile 
was blue shifted. This magnitude limit is 
comparable to the luminosity limit of stars showing emission, however stars with a blue shifted H$\alpha$ core did not
always show emission wings. The core velocity evidently is a more sensitive diagnostic of outflows. 
The amplitude of $v_{bis}$ correlates with the luminosity (Figure 10). Where AGB stars are well separated in color from
the RGB stars, near $log L/L_{\odot}$=2.3 to 2.6, the AGB stars exhibit larger values of $v_{bis}$ and more variability than RGB stars
(Figure 11). At higher luminosities, where the AGB and RGB objects can not be distinguished from one another in the CMD, the
bisector velocities are all significantly higher than stars at fainter magnitudes on the RGB.

\subsection{\ion{Ca}{2} K Profiles}

Spectra of 53 red giant stars in the \ion{Ca}{2}~H$\&$K region were obtained in 2005 and the profiles of the 
Ca K core ($\lambda$3933) are shown in 
Figures 12 and 13 for most stars showing emission. Two other stars, K702 and K1029, not shown in these figures, exhibit Ca
emission too. The spectra had low signal to noise ratio (S/N $\approx$ 15~$-$~20 in the 
continuum and $\approx$ 5~$-$~10 in the core) and it is difficult to identify emissions in many cases. 
Continuum normalization is challenging in 
this spectral region because spectral synthesis demonstrates that hundreds of absorbing lines generally depress the
continuum substantially. A low order Chebyschev function was used to fit and normalize the local 
continuum away from the strong \ion{Ca}{2} lines. In many stars,
the noise may be comparable to the emission in the core of the K line, preventing 
measurement of the radial velocity of the central self-absorption in almost all spectra.
The presence of emission is determined by eye by comparison to the synthesized 
spectrum of K341 constructed with the Kurucz code \citep{kurucz01} using the physical parameters described in Section 3.2. 
This model spectrum contains only photospheric lines and no chromosphere was included in the model, 
thus making it excellent for detection of emission. It is clear that the stellar spectrum in the line core does 
not reach zero flux as would be expected in such a deep photospheric line. Some additional counts in the core may come
from the inaccurate sky subtraction. 
We found fourteen stars out of 53 stars where \ion{Ca}{2}~K emission is definite or highly likely. 
Eleven of these stars showed 
H$\alpha$ emission when measured one day earlier in 2005. In the 11 stars where the Ca asymmetry is obvious, 
the ratio of the blue to the red side of the Ca~K emission core
(the core asymmetry) could be assessed (see Table 8) and in 6 out of the 
11 cases, the core asymmetry differed from the  asymmetry as noted in H$\alpha$. 

\section{Discussion}

\subsection{The H$\alpha$ line}

One hundred and ten stars were observed during 2005 and 2006 and twenty-nine of these had H$\alpha$ emissions. The magnitude
of the faintest star showing H$\alpha$ emission above the continuum varied among our 4 observations. 
This is not surprising based on the variability in the presence,
strength, and emission asymmetries found by us and by others \citep{cacciari02}. The spectra of 2005 May 22 reveal emission in the
faintest giant, K582, at $log (L/L_{\odot})=2.36$ (V=14.48). This star and several others (K158, K260, K482, K875, K979)
displaying emission may be located
on the AGB as judged by the CMD (Figure 7) where they are distinct from the red giant branch. Stars brighter than V~$\sim$~14
can not be clearly separated into AGB and RGB from the color-magnitude diagram alone. 

Inspection of the bisector velocities reveals differences among the red giant stars. None of the stars brighter than 
V~$\sim$~14.5, $log (L/L_{\odot})>2.4-2.5$ have red-shifted H$\alpha$ cores (Figure 10); all are blue-shifted or 
exhibit no shift at all. This luminosity cutoff corresponds also to the limit of the H$\alpha$ emission wings which 
suggests 
that the blue-shift and the emission are related. The emission wings arise deep in the chromosphere as models have shown 
\citep{dupree01, mauas01}, and the dynamics of the upper chromosphere are reflected in the line core. Motions start in the
lower atmospheric layers and then progress outward through the chromosphere. The amount of the bisector shift increases with 
stellar luminosity (Figure 10). 

There are several stars that show faster outflow asymmetries than the generally low-speed outflows near 
$log L/L_{\odot}$=2.3 to 2.6 (Figure 10). 
These stars, K158, K260, K582, K875, and K979 display bisector velocities ranging from $-$6.7~km~s$^{-1}$ to 
$-$13.2~km~s$^{-1}$ in comparison to the remainder of stars in that magnitude interval where bisector velocities are typically
less than $-$5~km~s$^{-1}$. One of these stars, K260 showed the largest change in core-shift velocity measuring  
$-$3.0$\pm$1.2~km~s$^{-1}$ in 2005 and $-$10.6$\pm$0.9~km~s$^{-1}$ in 2006. The position of the high-outflow stars in the CMD
suggests they are AGB stars since they lie blueward of the fiducial AGB in Figure 7. These objects appear to be AGB stars, and
the relatively high outflows mark the presence of a substantial stellar wind. However, as Figure 11 shows, other stars near
the fiducial AGB do not have high outflow velocities. If these low velocity objects are also AGB stars, 
then this argues for an episodic outflow. 

The asymmetry of the H$\alpha$ wings indicates that most of the giant stars have inflow motions in the region where the wings
are formed. Since it appears likely that these stars are pulsating \citep{mayor01}, although this may be controversial 
for the metal deficient field giants \citep{carney01}, it is of interest to compare the asymmetry pattern with
that of Cepheids which are known pulsators. One well-studied Cepheid, $\ell$ Car (HD 84810), shows variable emission wings in 
H$\alpha$ similar to those found here \citep{baldry01}. The appearance of $B/R < 1$ asymmetry generally coincides with 
blue-shifted photospheric metal absorption lines; and the converse
applies, when $B/R > 1$, the photospheric lines are red-shifted. So a dynamic linking clearly exists between the 
photosphere and the regions forming the H$\alpha$ line wings. The radial
velocity of this long-period (35 days) Cepheid shows photospheric red-shifts for about half of its pulsation period, and
`inflow' H$\alpha$ line asymmetries for about 0.35 of its period. Inspection of radial velocity curves from metallic lines 
shows that longer-period Cepheids have red shifts for a greater proportion of their pulsation period \citep{nardetto01, petterson01}. 
Thus it is not surprising that the H$\alpha$ profiles in the red giants in M15 show a dominant inflow asymmetry. 

Models for dust-free Mira stars \citep{struck01} indicate that lower levels of the atmosphere can support radial pulsations 
which develop
into a steady outflow at larger distances. A similar behavior is suggested by the outflows detected in the cores of the 
H$\alpha$ lines for the M15 giants. Models of the H$\alpha$ line profiles for metal deficient giants show that the core is
formed at a mass column density substantially above the region forming the wings \citep{dupree01, mauas01}. 
Such models for red giants are needed to explore the dynamics of the atmospheres and to evaluate the mass loss rate. 

\subsection{\ion{Ca}{2} Emission}

Fourteen of the red giants showed emission in \ion{Ca}{2}~K and 12 spectra are shown in Figures 12 and 13. The lower 
luminosity limit is at least the same as found for the presence of H$\alpha$ emission, namely $log (L/L_{\odot})=2.36$. 
The luminosity limits of \ion{Ca}{2} $K_{2}$ emission and H$\alpha$ appear to be related. Calcium emission may well extend to 
fainter limits; our spectra of fainter stars did not have sufficient signal to identify emission in the center of the deep 
photospheric \ion{Ca}{2} line. However, three of the stars
displaying \ion{Ca}{2} K emission do not have H$\alpha$ emission, but this is not suprising since the presence of H$\alpha$ 
emission is known to vary (see Table 4). While the spectra are noisy, the \ion{Ca}{2} K asymmetries seem to include 
all possibilities: $B < R$, $B=R$, $B>R$. These asymmetries differ in 6 stars from the asymmetries of the H$\alpha$ wing 
in each star. Such a difference is not unexpected 
since the regions of formation of the Ca~K core and the H$\alpha$ wings are separated in the atmosphere
of a giant (Ca~$K_{2}$ emission forms higher in the atmosphere than H$\alpha$ wings); 
additionally H$\alpha$ shows variations in asymmetries over a few days \citep{cacciari02} which could contribute to 
the differences.

\subsection{Comparison with Other Studies}

A detailed spectroscopic study \citep{cacciari01} was made of 137 RGB stars in NGC~2808 which extended to $\sim$ 3 magnitudes
below the tip of the red giant branch. The majority of their targets were at lower resolution than we have here, however 
20 were sampled at high resolution. 
Emission in H$\alpha$ was detected down to a limit of $log (L/L_{\odot})=2.5$ which
is about 0.15 magnitude brighter than we find in M15. NGC~2808 is more metal rich ([Fe/H]=$-$1.15, \citet{harris01}) than M15 
([Fe/H]=$-$2.26, \citet{harris01}) which may account for the slight difference. 
On the other hand, the distance modulus for
M15 was recently determined to be 15.53$\pm$0.21 using the zero-age horizontal branch level as a 
distance indicator \citep{cho01} which would bring the luminosities into closer agreement. 
In the \citet{cacciari01} spectra taken at highest resolution, the red giants in NGC~2808 have wing emission in H$\alpha$ 
indicative of inflow ($B/R>1$) in the majority of stars, similar to our results.
Other surveys generally contained only the brightest stars in the clusters, and their luminosity limits extend only to 
$log (L/L_{\odot})=2.7$ \citep{bates02, lyons01}. \citet{smith01} noted H$\alpha$ emissions in metal-poor field giants
brighter than $log (L/L_{\odot})=2.5$. 

While the luminosity limits of the H$\alpha$ emission are similar in M15 and NGC~2808, the distribution of the emission 
with luminosity and effective temperatures differs between the clusters. Figure 14 compares the fraction of stars with 
H$\alpha$ emission in M15 from this study with NGC~2808 \citep{cacciari01}. At the same luminosity, the M15 giants exhibit a 
lower percentage of H$\alpha$ emissions than is found in NGC~2808. However, at the same effective temperature, the 
fraction of stars showing emission is the opposite; fewer stars have emission at the same effective temperature in NGC~2808 
than in M15. Although the fraction of stars with emission generally increases with luminosity, the differences result
principally because the CMD for NGC~2808 (which is more metal rich than M15) lies at lower effective temperatures at the same
luminosity. Assessing the emission fraction as a function of stellar radius (Figure 14, \textit{right panel}) suggests that 
the fractions are comparable except at large values of the stellar radius. Possibly the coolest stars can not support the 
thick chromosphere necessary to produce emission \citep{dupree01, mauas02} and/or the pulsational characteristics of the 
atmospheres differ.

\citet{cacciari01} were able to measure significant core shifts of H$\alpha$ in 7 stars of their sample of giants 
in NGC~2808 observed with the high resolution spectrograph UVES (20 stars). 
Outflow velocities more negative than $-$2~km~s$^{-1}$ appear for stars brighter than $log (L/L_{\odot})=3.16$
but there is little velocity data for the fainter giants. More stars had measurable velocities in the Na~D lines, and 
outflows from 1$-$4~km~s$^{-1}$ became apparent at $log (L/L_{\odot})=3.1$ and brighter. M4, another cluster 
of similar metallicity as NGC~2808 did not have coreshifts (more negative than $-$2~km~s$^{-1}$) 
either in H$\alpha$ or Na~D in any of $\approx$10 stars that have luminosities $log (L/L_{\odot})<3.3$ 
\citep{kemp01}. By contrast, the velocities in M15 indicated systematic outflow (more negative than $-$2~km~s$^{-1}$) 
in the H-alpha core occur at lower luminosities, $log (L/L_{\odot})=2.5$. Thus there are possible signs of a metal dependency 
in the outflow with higher velocities in metal poor objects. More complete sampling of outflows in other clusters
is needed.

The luminosity limit \citep{cacciari01} for Ca K emission lines was 
$log (L/L_{\odot})\sim2.6$ which is comparable to the H$\alpha$ limit in NGC~2808 and similar to the results for M15. However
our data did not have sufficient signal to noise to check the fainter stars. In sum, while the chromospheric emissions appear 
independent of metallicity, outflows may be enhanced in low metallicity stars.

\subsection{\textit{Spitzer} Stars}

\citet{boyer01} have observed M15 with the \textit{Spitzer} Space Telescope using the IRAC and MIPS instruments. 
They concluded that a significant 
amount of dust ($9 \ \pm 2 \times 10^{-4} M_{\odot}$) occurs near the center of the cluster and suggested that this dust 
comes from the mass-loss of the brightest giant stars. 
Twenty-three stars were identified as dusty IR sources and 
their IRAC colors indicate that these are AGB stars. We observed 12 of these stars (Table 1) and could
confirm their cluster membership with radial velocity measurements (Table 5). We noted one source, designated by 
\citet{boyer01} as unidentified (SSTU J212953.33 +120910.7) is associated with K272 at the same coordinates.
All of the observed \textit{Spitzer} stars showed blue shifts in H$\alpha$. Six of them (K224, K421, K479, K672, K709 and 
GEB254) have strong H$\alpha$ emission, but only one (K479) had outflow asymmetry in H$\alpha$ (B$<$R). 
Other stars in the \textit{Spitzer} field (K144, K260, K393, K431, K447, K462, K702) were not identified as dusty stars, 
yet their H$\alpha$ line emissions and blueshifts are similar to those stars identified by \textit{Spitzer} observations 
as having an IR excess. Evidently not all stars produce dusty shells. The H$\alpha$ emission profiles observed here do not
seem to be related to a prior phase when the star produced material that cooled down to produce an IR excess.

All twelve \textit{Spitzer} stars have the same colors and luminosities as other RGB stars 
on the CMD (Figure 7) and their bisector velocities also show similar values (Figure 10) as other RGB stars at the same 
luminosity. However the bisector velocity does not appear to correlate with any IRAC colors or magnitudes.
The intensity ratio of short wavelength and long wavelength emission peaks (B/R) and the strength of the H$\alpha$ 
emission wings are very similar to other stars in this color and 
luminosity region of the CMD. In this region, where the \textit{Spitzer} sources are located, the differences between 
the AGB stars and RGB stars are hard to discern spectroscopically.

\textit{Spitzer} showed \citep{boyer01} that only some stars in this region of the CMD have a mid$-$IR excess. 
Assuming that the \textit{Spitzer} sources must have had strong stellar winds to produce dust and the 
current H$\alpha$ emission profiles are not related to the episode of dust production, one can
conclude that the mass loss is not continuous. \citet{origlia01} used ISOCAM images to study red giants in
globular clusters but the large pixel size of ISOCAM made it difficult to identify stellar sources. 
Frequently, several stars were candidates for each mid-IR source in a 15 arcsec square area, and the brightest one 
was selected. They concluded that strong mass loss occurred only among RGB stars located at the red giant tip 
[$log (L/L_{\odot})>=3$]. Not all of the luminous stars identified in this way had in IR excess
indicating dust, suggesting that the mass loss was episodic. The results reported here have no ambiguity in 
identification, and demonstrate the presence of episodic mass loss over a much greater extent in luminosity. 
These stars must have passed through several active phases with very strong stellar winds during their 
lifetimes on the AGB. 

\section{Conclusions}

Differences found in the profiles of the H$\alpha$ line 
give insight into the atmospheric structure and dynamics. Stars which are physically larger show more emission. At
a given luminosity, the M15 stars exhibit a consistently lower {\it fraction} of stars with emission, than found
in NGC~2808, a cluster which is more metal rich. This suggests the M15 chromospheres may have less material 
than stars at the same luminosity in NGC~2808 as demonstrated by detailed modeling of emission wings and 
consistent with the decrease in radius with metallicity. 

Stars in M15 have H$\alpha$ emission wings that vary in time so that the magnitude of the faintest giant showing 
emission changes among the different dates of observation. The lower limit to the presence of H$\alpha$ emission
in M15 [$log (L/L_{\odot})=2.36$] is comparable to that found in NGC~2808 \citep{cacciari01}.

M15 exhibits continuous outflows at lower luminosities and with higher velocities than the more metal rich clusters, 
NGC~2808 and M4, hinting at a metallicity dependence. These outflows may decrease the chromospheric material and account for
the lower fraction of stars with emission wings in M15. Detailed modeling is necessary to 
evaluate the mass loss rate from the line profiles. To identify mass loss as the solution to the second parameter problem 
will also require more complete measurements of clusters with varying metallicities. 

The bisector velocity of the H$\alpha$ core along the RGB indicates outflow (negative
velocities) or no motion for stars brighter than $log (L/L_{\odot})=2.5$; the
outflow velocity increases with increasing stellar luminosity. However, AGB stars near
$log (L/L_{\odot})\sim2.4$ have bisector velocities comparable in value to those at the 
tip of the RGB and also exhibit larger changes in velocity between observation 
than the RGB stars. We take this as evidence of more substantial and episodic mass outflow on the AGB.

\ion{Ca}{2}~K emission is detected at least to the limits of H$\alpha$ emission; 
however, the asymmetry in the \ion{Ca}{2}~K core, where measurable, may differ from the asymmetry 
measured in the H-alpha wings perhaps due to time variability or different
line-forming regions.

Twelve stars identified in Spitzer observations as dusty IR sources and AGB 
stars \citep{boyer01} have radial velocities consistent with cluster
membership. The similarities in H$\alpha$ line profile characteristics 
between the \textit{Spitzer} sources and other red giants in M15 suggests the IR 
emission attributed to circumstellar dust must be produced by an episodic 
process.  

\acknowledgments{Observations reported here were obtained at the MMT Observatory, a joint 
facility of the Smithsonian Institution and the University of Arizona. 
We are grateful to the scientists at CfA who are developing and characterizing Hectochelle: Nelson Caldwell,
Daniel G. Fabricant, Gabor Furesz, and David W. Latham. The authors also would like to thank John
Roll and Maureen A. Conroy for developing SPICE software, Michael Calkins and Perry Berlind for their help during the
observations, Robert Kurucz for the spectrum synthesis, and Martha Boyer for helpful comments on the manuscript. Kyle
Cudworth kindly provided coordinates and photometry for M15 stars. 
Szabolcs~Meszaros is supported in part by a SAO Predoctoral Fellowship, NASA, and the Hungarian OTKA Grant TS049872.
This research is also supported in part by the Smithsonian Astrophysical Observatory.}

\clearpage

\begin{figure}
\includegraphics[width=3in,angle=0]{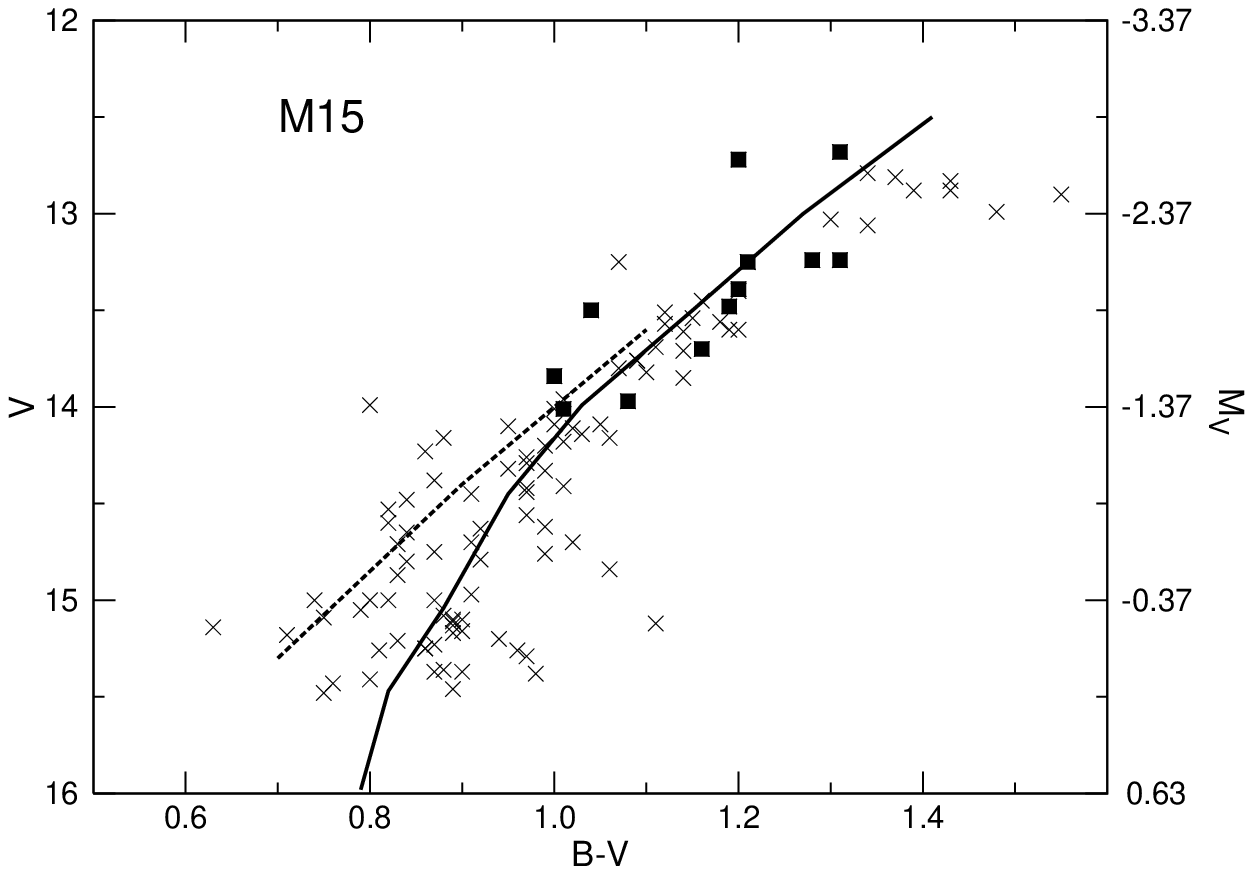}
\caption{Color-magnitude diagram for all stars observed in 2005 and 2006. 
The solid line shows the fiducial curve of the RGB and the dashed line shows the fiducial curve of the 
AGB for M15 taken from observations of \citet{durrell01}. Dusty giants
identified with \textit{Spitzer} Space Telescope \citep{boyer01} and observed with Hectochelle, are denoted by 
squares. The redward "hook" seen among the brightest stars in M15 is not intrinsic to 
the stars but rather results from saturation of the photographic images \citep{sneden01}. Absolute magnitudes are obtained by
assuming $(m-M)_V=15.37$ \citep{harris01}.}
\end{figure}

\begin{figure}
\includegraphics[width=3in,angle=0]{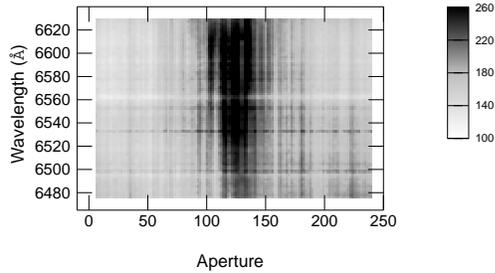}
\caption{Variations in the intensity of the sky apertures in the H$\alpha$ filter. Dark area indicates the highest count
level. See scale at right. An anomalous intensity pattern occurs, which is currently not understood. Special extraction
patterns were used for the sky fibers (see text).}
\end{figure}

\begin{figure}
\includegraphics[width=3in,angle=0]{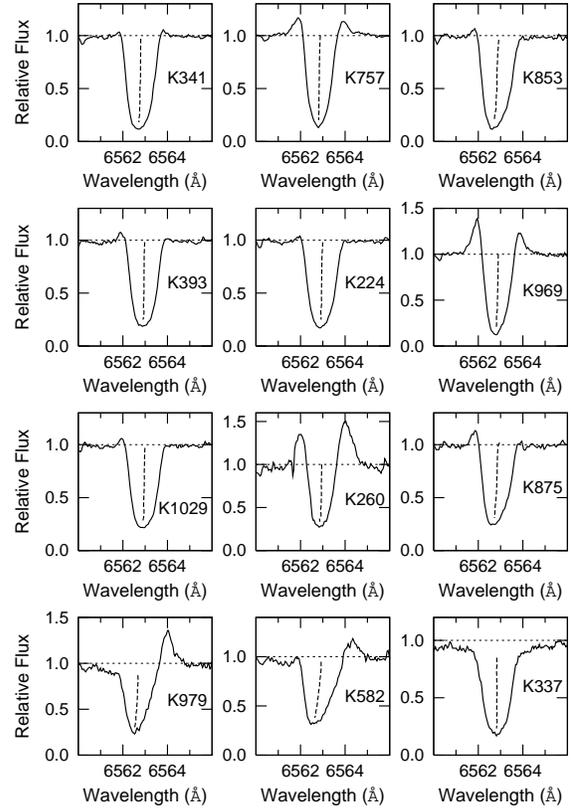}
\caption{Normalized spectra of red giants in M15 which showed emission in H$\alpha$ on 2005 May 22. The dashed line marks
the bisector.
The spectra are arranged in order of decreasing V magnitude; the brightest is at the top left and the stars 
become fainter from left to right. The wavelength scale is corrected for heliocentric velocity. 
K337 is an example of an H$\alpha$ profile without emission. Stars K260 (Figures 3 and 4), K341 (Figures 3$-$6), K757 
(Figures 3 and 4), and K969 (Figures 3 and 6) showed large variations in H$\alpha$ emission during the observation period.}
\end{figure}

\begin{figure}
\includegraphics[width=3in,angle=0]{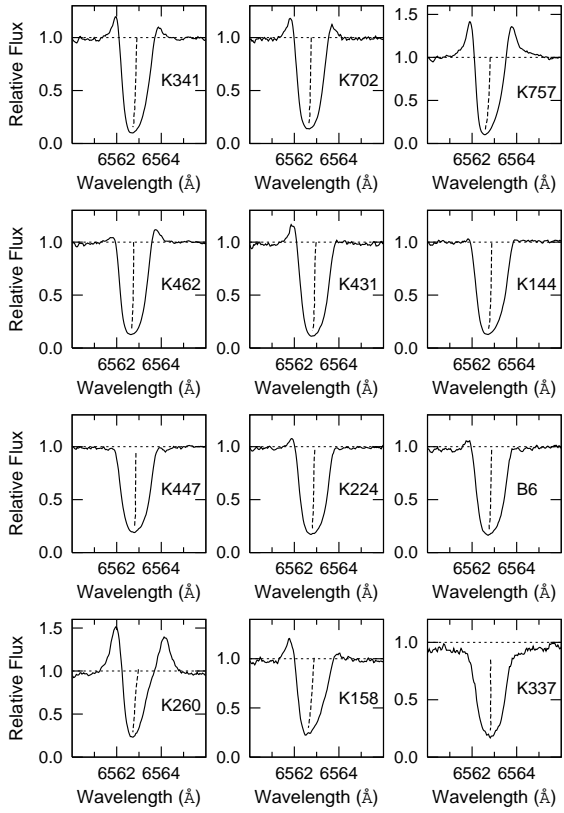}
\caption{Normalized spectra of red giants in M15 which showed emission in H$\alpha$ on 2006 May 11. For explanation please
see Figure 3. Stars K260 (Figures 3 and 4), K341 (Figures 3$-$6), K431 (Figures 4 and 5), and K757 
(Figures 3 and 4) showed large variations in H$\alpha$ emission.}
\end{figure}

\begin{figure}
\includegraphics[width=3in,angle=0]{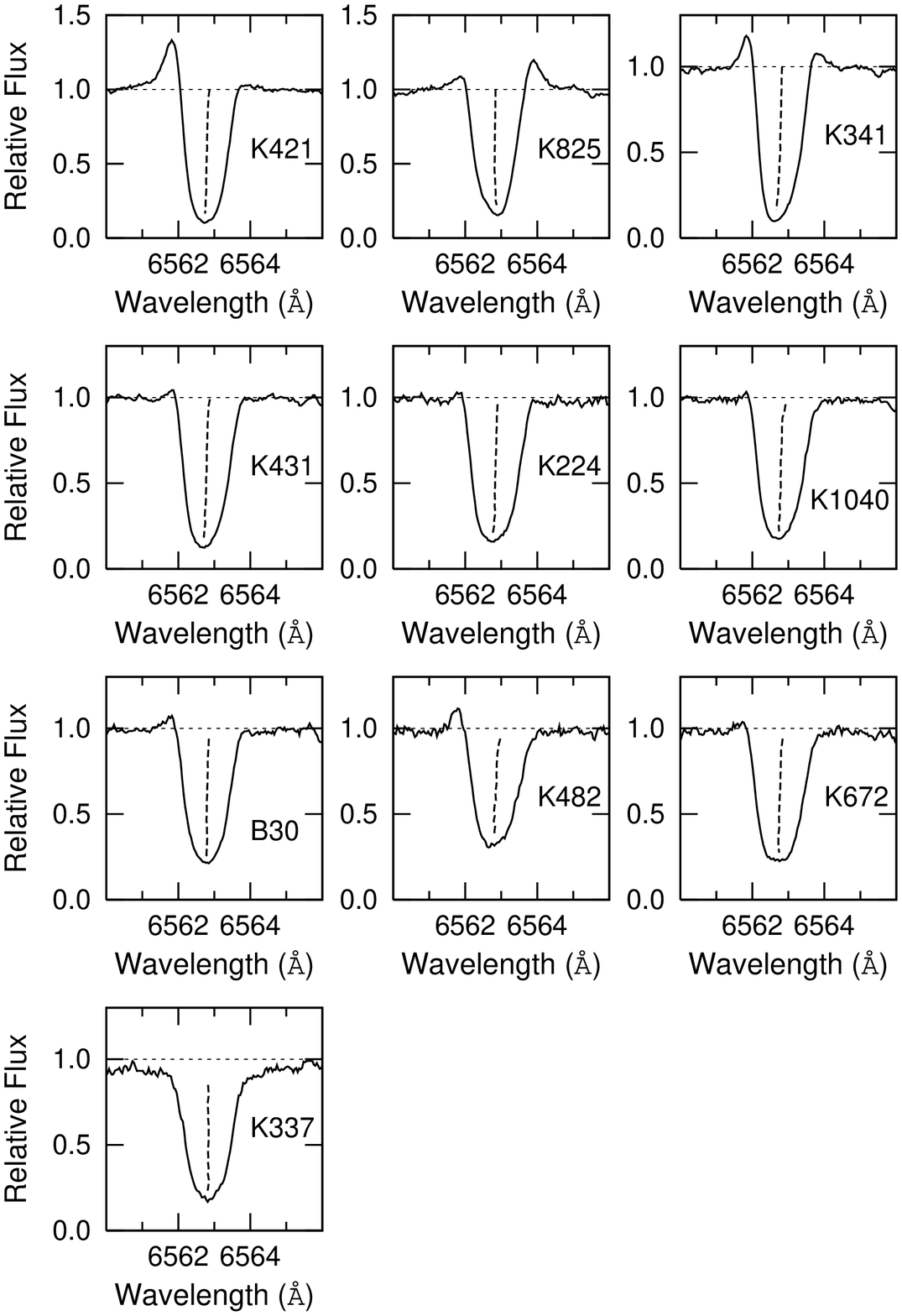}
\caption{Normalized spectra of red giants in M15 which showed emission in H$\alpha$ on 2006 October 4. For explanation please
see Figure 3. Stars K341 (Figures 3$-$6), K431 (Figures 4 and 5) showed large variations in H$\alpha$ emission.}
\end{figure}

\begin{figure}
\includegraphics[width=3in,angle=0]{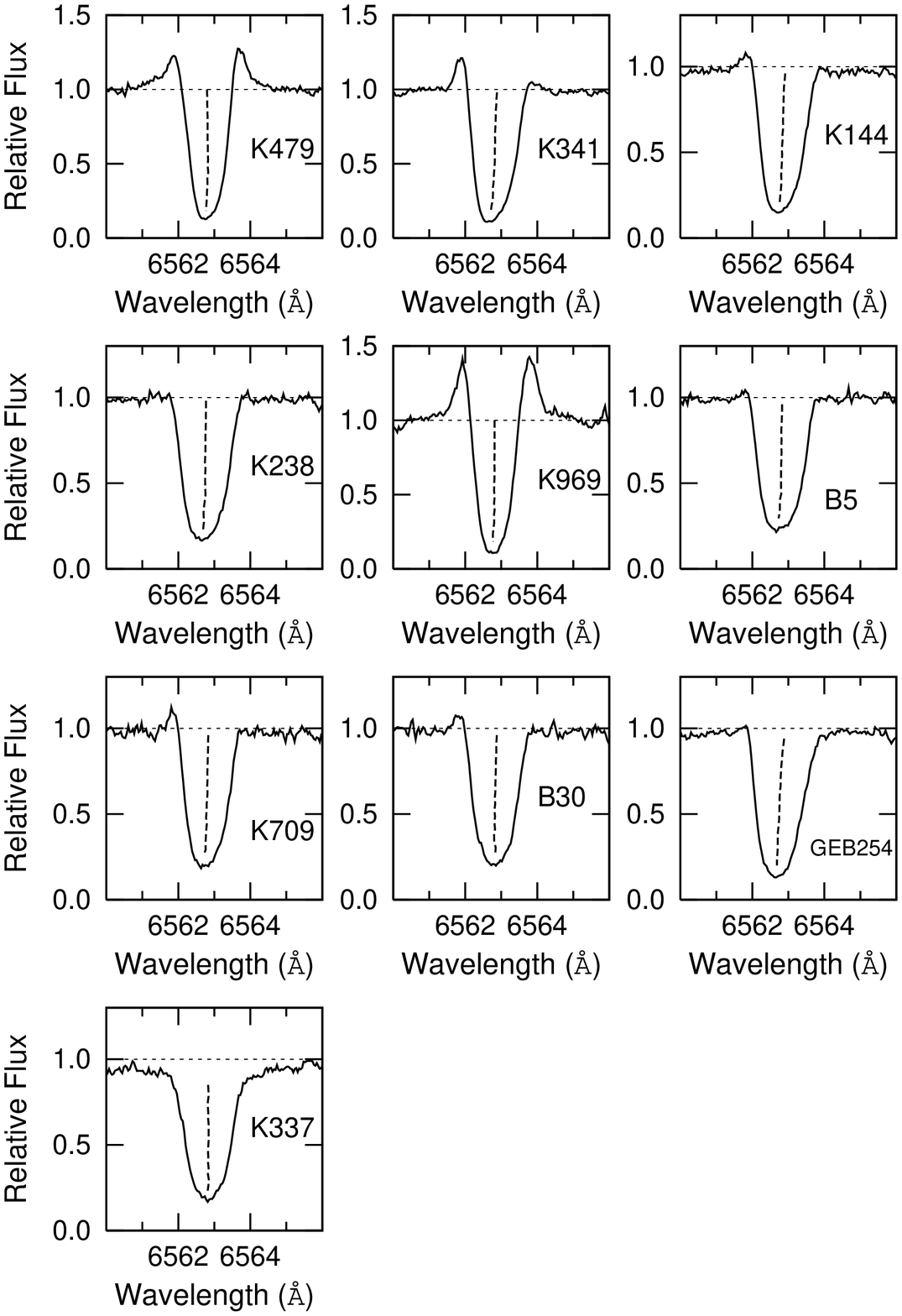}
\caption{Normalized spectra of red giants in M15 which showed emission in H$\alpha$ on 2006 October 7. For explanation please
see Figure 3. Stars K341 (Figures 3$-$6), K969 (Figures 3 and 6) showed large variations in H$\alpha$ emission.}
\end{figure}

\begin{figure}
\includegraphics[width=3in,angle=0]{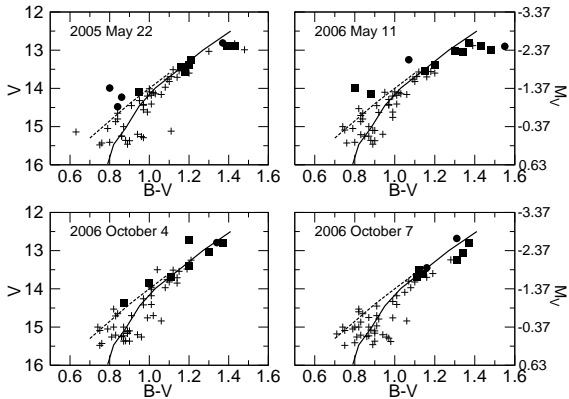}
\caption{Color-magnitude diagrams for all M15 stars observed in 2005 and 2006. 
Stars with $H\alpha$ emission and with B$<$R (indicating outflow) are marked with circles; 
stars with B$>$R emission wings (suggests inflow) are denoted by squares. 
The solid line shows the fiducial curve of the RGB; dashed lines show the fiducial curve of the 
AGB for M15 from observations of \citet{durrell01}.}
\end{figure}

\begin{figure}
\includegraphics[width=3in,angle=0]{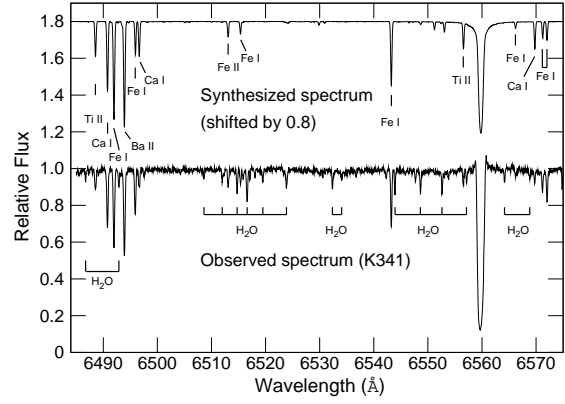}
\caption{Kurucz synthesized spectrum shifted by 0.8 in relative flux and corrected for the star's radial velocity
shown above the observed spectrum of K341.
Atmospheric water vapor and other elements are marked.}
\end{figure}

\begin{figure}
\includegraphics[width=3in,angle=0]{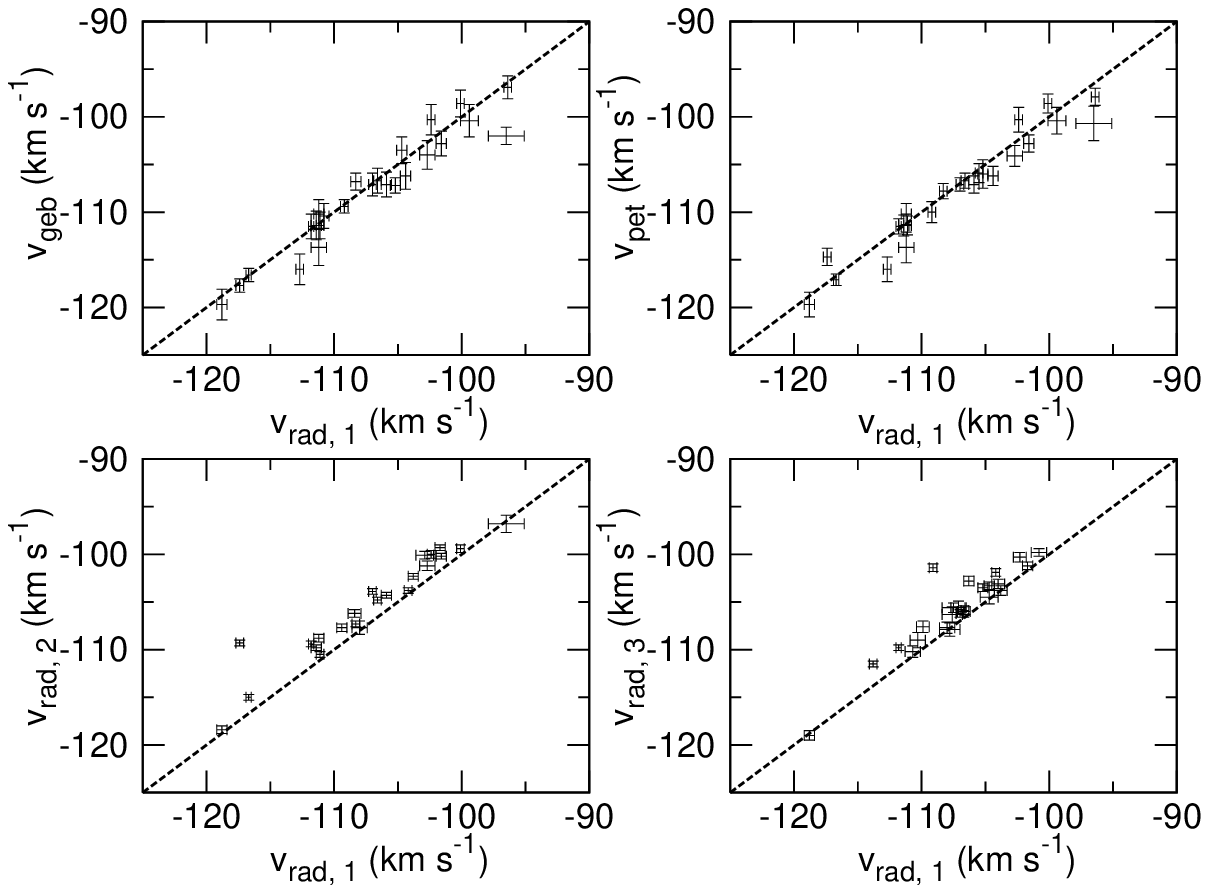}
\caption{Top left: Radial velocities measured in this paper on 2005 May 22 ($v_{rad,1}$) for the same stars 
observed by \citet{gebhardt01} ($v_{geb}$). 
Top right: Radial velocities measured in this paper on 2005 May 22 ($v_{rad,1}$)
for the same stars observed by \citet{peterson04} ($v_{pet}$). There is good agreement between
observations taken on 2005 May 22 and observations for the same stars from \citet{gebhardt01} and \citet{peterson04}. 
Lower left: Radial velocity measured with Hectochelle for the same stars observed on 2005 
May 22 ($v_{rad,1}$) compared to 2006 May 11 ($v_{rad,2}$). 
The velocity offset between 2006 May 11 and 2005 May 22 is $+1.9 \ \pm 0.5$~km~s$^{-1}$. 
Lower right: Radial velocities for the same stars measured with Hectochelle on 2005 May 22 ($v_{rad,1}$) compared to 
2006 October 4 ($v_{rad,3}$). The velocity offset between 2006 October 4 and 2005 May 22 is $+0.9 \ \pm 0.5$~km~s$^{-1}$. 
The dashed line marks a 1:1 relation. The offsets are
applied to our radial velocities for the 2006 May and 2006 October spectra (see text).}
\end{figure}

\begin{figure}
\includegraphics[width=3in,angle=0]{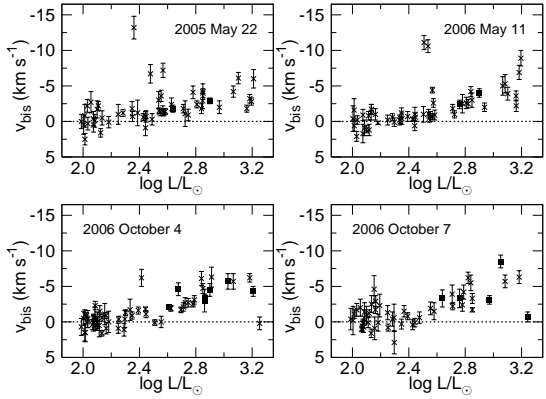}
\caption{The velocity difference ($v_{bis}$) between the top and the bottom of the
bisector of H$\alpha$ as a function of luminosity. Negative values indicate a blueshifted core (outward motion), 
positive values denote a red shifted core (inward motion). A predominant outward motion sets in near 
$log L/L_{\odot}$ $\approx$ 2.4 
and increases in velocity towards more luminous stars. Dusty giants
identified with \textit{Spitzer} Space Telescope \citep{boyer01} and observed with Hectochelle, are denoted by 
squares. See text for discussion of the outlying stars between $log L/L_{\odot}$=2.3 and 2.6 with velocities 
more negative than $-10$~km~s$^{-1}$ which appear to be AGB stars.}
\end{figure}

\begin{figure}
\includegraphics[width=3in,angle=0]{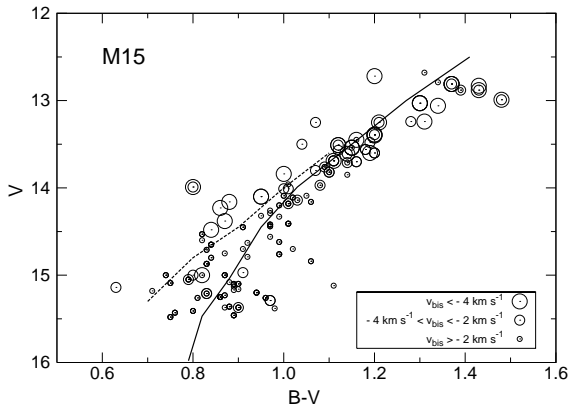}
\caption{Color-magnitude diagram for all M15 stars observed in 2005 and 2006, where the size of the circle indicates the
velocity of the H$\alpha$ bisector asymmetry. Big circle: $v_{bis} 
<-4$~km~s$^{-1}$, medium circle: $-4$ km s$^{-1} < v_{bis} < -2$~km~s$^{-1}$, small circle: $v_{bis} > -2$~km~s$^{-1}$.
Concentric circles generally indicate multiple observations of the same star.}
\end{figure}

\begin{figure}
\includegraphics[width=3in,angle=0]{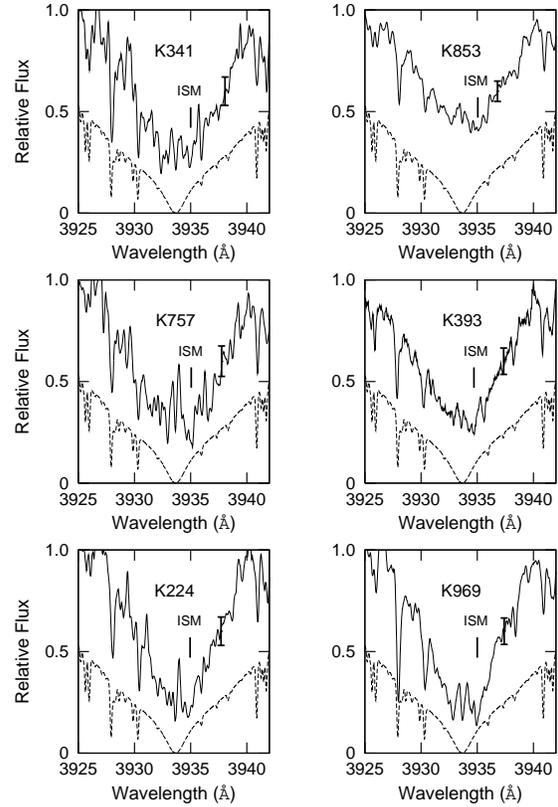}
\caption{Spectra of some red giants in M15 which showed emission in \ion{Ca}{2}~K. The observed spectra are shifted up by 0.1. 
The Kurucz model of K341 is denoted by a dashed line. The spectra are arranged in order of decreasing V magnitude; the 
brightest is at the top left and the stars become fainter from left to right. The spectra are smoothed to make the 
spectral features more visible. Error bars show the photon noise in the original, unsmoothed spectra. The line marked ISM
denotes absorption by the interstellar medium.}
\end{figure}

\begin{figure}
\includegraphics[width=3in,angle=0]{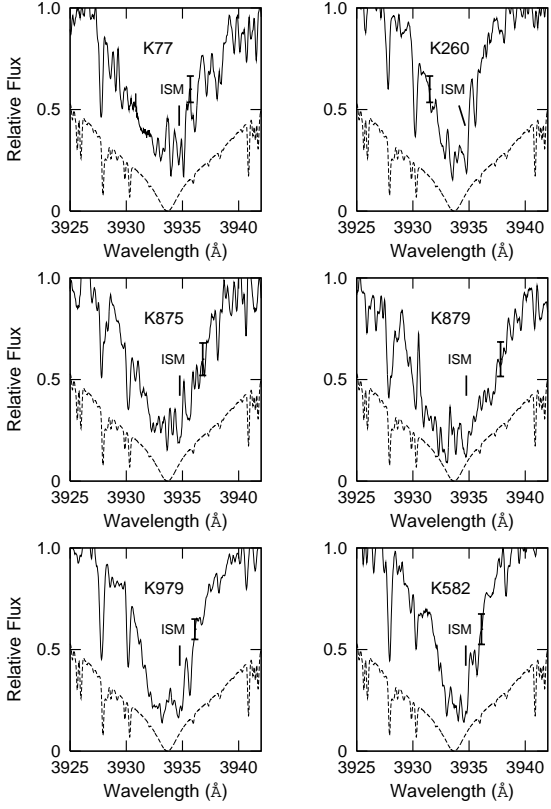}
\caption{Spectra of some red giants in M15 which showed emission in \ion{Ca}{2}~K. For explanation please
see Figure 12.}
\end{figure}

\begin{figure}
\includegraphics[width=3in,angle=0]{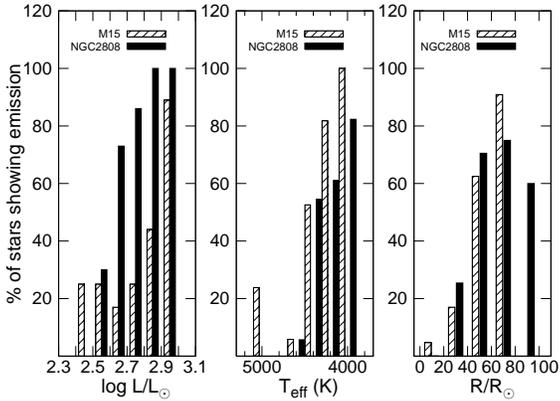}
\caption{Left panel: Percentage of stars showing emission in M15 and NGC~2808 as a function of luminosity. Values for the 
$log (L/L_{\odot})=2.9$ bin include all brighter stars as well. Center panel: 
Percentage of stars showing emission in M15 and NGC~2808 as a function of effective temperature. The lowest and highest bin
includes the lower and upper limits respectively. Right panel: 
Percentage of stars showing emission in M15 and NGC~2808 as a function of stellar radius.}
\end{figure}

\clearpage

\LongTables
\begin{deluxetable*}{lcccccccccc}
\tablecolumns{11}
\tabletypesize{\scriptsize}
\tablecaption{Photometric Data of Observed Cluster Members}
\tablewidth{0pt}
\tablehead{
\colhead{ID No. \tablenotemark{a}}           & \colhead{$RA(2000)$ \tablenotemark{b} }    &
\colhead{Dec(2000) \tablenotemark{b} }          & \colhead{B}  &
\colhead{V}          & \colhead{J}    &
\colhead{H}  & \colhead{K}  &
\colhead{B$-$V} & \colhead{V$-$K} & \colhead{Obs. \tablenotemark{c} }
}
\startdata
B5 \tablenotemark{d} & 21 29 08.43 & +12 09 11.8 & 14.63 & 13.51 & 11.368 & 10.726 & 10.645 & 1.12 & 2.87 & 1,3,4 \\
B6 \tablenotemark{d} & 21 29 12.36 & +12 10 49.8 & 14.69 & 13.54 & 11.411 & 10.775 & 10.656 & 1.15 & 2.88 & 1,2,3,4 \\
B16 \tablenotemark{d} & 21 29 53.12 & +12 12 31.1 & 14.85 & 13.76 & 11.732 & 11.154 & 11.029 & 1.09 & 2.73 & 3 \\
B30 \tablenotemark{d} & 21 30 44.13 & +12 11 22.6 & 14.80 & 13.69 & 11.496 & 10.927 & 10.807 & 1.11 & 2.88 & 1,2,3,4 \\
C3 \tablenotemark{e} & 21 29 13.06 & +12 11 15.0 & 15.49 & 14.65 & 12.698 & 12.156 & 12.039 & 0.84 & 2.61 & 1,2,3,4 \\
C20 \tablenotemark{e} & 21 29 52.32 & +12 19 39.7 & 16.35 & 15.46 & 13.646 & 13.089 & 12.991 & 0.89 & 2.47 & 1,2,3,4 \\
C35 \tablenotemark{e} & 21 30 49.28 & +12 07 31.2 & 15.77 & 15.14 & 13.398 & 12.965 & 12.855 & 0.63 & 2.29 & 1 \\
GEB 254 \tablenotemark{f} & 21 29 58.15 & +12 09 46.7 & 14.55 \tablenotemark{g} & 13.24 \tablenotemark{g} & 9.560 & 9.964 & 9.777 & 1.31 & 3.46 & 4,5 \\
GEB 289 \tablenotemark{f} & 21 29 59.37 & +12 10 02.9 & 14.54 \tablenotemark{g} & 13.50 \tablenotemark{g} & 11.079 & 10.449 & 10.311 & 1.04 & 3.12 & 3,5 \\
K12 & 21 29 30.77 & +12 06 32.7 & 15.84 & 15.05 & 13.386 & 12.886 & 12.851 & 0.79 & 2.20 & 1,2,3,4 \\
K21 & 21 29 33.12 & +12 12 51.0 & 15.74 & 15.00 & 13.323 & 12.834 & 12.726 & 0.74 & 2.27 & 2,3,4 \\
K22 & 21 29 33.53 & +12 04 55.3 & 15.42 & 14.41 & 12.482 & 11.938 & 11.809 & 1.01 & 2.60 & 1,2,3,4 \\
K26 & 21 29 34.62 & +12 03 19.2 & 15.87 & 15.00 & 13.197 & 12.693 & 12.566 & 0.87 & 2.43 & 1,2,3,4 \\
K27 & 21 29 35.08 & +12 06 03.7 & 15.84 & 15.09 & 13.451 & 12.994 & 12.894 & 0.75 & 2.20 & 2,3,4 \\
K31 & 21 29 35.87 & +12 08 27.3 & 16.04 & 15.21 & 13.529 & 13.006 & 12.882 & 0.83 & 2.33 & 2,3,4 \\
K42 & 21 29 38.00 & +12 11 58.2 & 16.10 & 15.23 & 13.452 & 12.925 & 12.887 & 0.87 & 2.34 & 3,4 \\
K47 & 21 29 38.72 & +12 11 53.0 & 15.13 & 14.11 & 12.152 & 11.588 & 11.488 & 1.02 & 2.62 & 1,2 \\
K56 & 21 29 40.04 & +12 16 00.2 & 16.27 & 15.37 & 13.651 & 13.107 & 12.966 & 0.90 & 2.40 & 1,2,3,4 \\
K60 & 21 29 41.25 & +12 07 19.4 & 16.07 & 15.26 & 13.532 & 12.991 & 12.928 & 0.81 & 2.33 & 3,4 \\
K64 & 21 29 42.96 & +12 09 53.4 & 16.11 & 15.25 & 13.504 & 13.003 & 12.884 & 0.86 & 2.37 & 1,2,3,4 \\
K69 & 21 29 43.77 & +12 08 33.3 & 15.36 & 14.45 & 12.501 & 11.969 & 11.854 & 0.91 & 2.60 & 1,2,4 \\
K70 & 21 29 43.59 & +12 15 47.4 & 15.27 & 14.32 & 12.338 & 11.746 & 11.653 & 0.95 & 2.67 & 1 \\
K77 & 21 29 44.65 & +12 07 30.8 & 14.92 & 13.82 & 11.752 & 11.148 & 11.026 & 1.10 & 2.79 & 1,2,3,4 \\
K87 & 21 29 45.81 & +12 08 45.5 & 14.87 & 13.80 & 11.776 & 11.225 & 11.075 & 1.07 & 2.73 & 4 \\
K89 & 21 29 46.07 & +12 11 31.5 & 15.35 & 14.53 & 12.564 & 12.055 & 11.955 & 0.82 & 2.58 & 2,3,4 \\
K92 & 21 29 46.70 & +12 03 20.7 & 16.14 & 15.20 & 13.359 & 12.848 & 12.791 & 0.94 & 2.41 & 1,2,3,4  \\
K105 & 21 29 47.39 & +12 09 04.7 & 16.26 & 15.29 & 13.477 & 13.023 & 12.940 & 0.97 & 2.35 & 1,4 \\
K112 & 21 29 47.78 & +12 11 30.7 & 15.96 & 15.08 & 13.167 & 12.654 & 12.592 & 0.88 & 2.49 & 4 \\
K114 & 21 29 47.87 & +12 08 45.4 & 14.99 & 13.85 & 11.781 & 11.245 & 11.080 & 1.14 & 2.77 & 3 \\
K129 & 21 29 48.63 & +12 11 45.8 & 15.23 & 14.26 & 12.292 & 11.725 & 11.620 & 0.97 & 2.64 & 3 \\
K133 & 21 29 48.84 & +12 10 25.3 & 15.99 & 15.10 & 13.435 & 13.014 & 12.898 & 0.89 & 2.20 & 2,3 \\
K136 & 21 29 49.13 & +12 09 03.7 & 15.80 & 15.00 & 13.217 & 12.811 & 12.707 & 0.80 & 2.29 & 4 \\
K137 & 21 29 49.45 & +12 08 27.0 & 16.36 & 15.38 & 13.628 & 13.134 & 13.059 & 0.98 & 2.32 & 3 \\
K144 & 21 29 49.79 & +12 11 05.9 & 14.40 & 13.06 & 10.745 & 10.053 & 9.944 & 1.34 & 3.12 & 2,4 \\
K145 & 21 29 49.80 & +12 12 29.9 & 16.24 & 15.36 & 13.557 & 13.090 & 12.957 & 0.88 & 2.40 & 1,2,3 \\
K146 & 21 29 49.94 & +12 08 05.3 & 14.69 & 13.57 & 11.427 & 10.843 & 10.726 & 1.12 & 2.84 & 2 \\
K151 & 21 29 50.14 & +12 07 52.2 & 16.00 & 15.11 & 13.303 & 12.757 & 12.699 & 0.89 & 2.61 & 3 \\
K152 & 21 29 50.16 & +12 06 40.7 & 16.11 & 15.25 & 13.553 & 12.999 & 12.877 & 0.86 & 2.37 & 1,2,3 \\
K153 & 21 29 49.92 & +12 18 12.1 & 16.23 & 15.48 & 13.684 & 13.195 & 13.067 & 0.75 & 2.41 & 1,3,4 \\
K158 & 21 29 50.29 & +12 09 02.8 & 15.04 & 14.16 & 12.380 & 11.879 & 11.806 & 0.88 & 2.35 & 2 \\
K202 & 21 29 51.89 & +12 06 38.9 & 16.06 & 15.17 & 13.420 & 12.906 & 12.838 & 0.89 & 2.33 & 3 \\
K224 & 21 29 52.31 & +12 10 51.5 & 14.59 & 13.39 & 11.235 & 10.672 & 10.515 & 1.20 & 2.88 & 1,2,3,5 \\
K238 & 21 29 52.65 & +12 10 44.0 & 14.52 & 13.24 & 11.041 & 10.429 & 10.305 & 1.28 & 2.94 & 4,5 \\
K255 & 21 29 53.12 & +12 12 31.1 & 14.85 & 13.76 & 11.732 & 11.154 & 11.029 & 1.09 & 2.73 & 1,2 \\
K260 & 21 29 53.31 & +12 09 34.0 & 14.79 & 13.99 & 12.486 & 12.125 & 12.053 & 0.80 & 1.94 & 1,2 \\
K272 & 21 29 53.57 & +12 09 10.7 & 14.67 & 13.48 & 11.307 & 10.736 & 10.606 & 1.19 & 2.87 & 3,5 \\
K288 & 21 29 53.79 & +12 10 20.3 & 14.85 & 13.71 & 11.667 & 11.120 & 10.966 & 1.14 & 2.74 & 1,3 \\
K328 & 21 29 54.73 & +12 08 59.2 & 14.86 & 13.70 & 11.591 & 11.065 & 10.906 & 1.16 & 2.80 & 2,4,5 \\
K337 & 21 29 55.05 & +12 02 48.5 & 15.75 & 14.76 & 12.863 & 12.311 & 12.200 & 0.99 & 2.56 & 2,3,4 \\
K341 & 21 29 54.93 & +12 13 22.5 & 14.18 & 12.81 & 10.455 & 9.796 & 9.695 & 1.37 & 3.12 & 1,2,3,4 \\
K361 & 21 29 55.29 & +12 09 13.7 & 14.97 & 13.96 & 12.029 & 11.475 & 11.364 & 1.01 & 2.60 & 3 \\
K393 & 21 29 55.73 & +12 11 33.8 & 14.46 & 13.25 & 11.042 & 10.452 & 10.315 & 1.21 & 2.94 & 1,2 \\
K421 & 21 29 56.18 & +12 10 17.9 & 13.92 & 12.72 & 10.414 & 9.781 & 9.649 & 1.20 & 3.07 & 3,5 \\
K431 & 21 29 56.18 & +12 12 33.8 & 14.33 & 13.03 & 10.759 & 10.144 & 10.039 & 1.30 & 2.99 & 1,2,3 \\
K447 & 21 29 56.45 & +12 10 29.4 & 14.32 & 13.25 & 11.132 & 10.613 & 10.468 & 1.07 & 2.78 & 2 \\
K462 & 21 29 56.67 & +12 09 46.3 & 14.45 & 12.90 & 10.534 & 9.860 & 9.722 & 1.55 & 3.18 & 2 \\
K476 & 21 29 56.72 & +12 13 10.5 & 15.55 & 14.63 & 12.714 & 12.202 & 12.130 & 0.92 & 2.50 & 2 \\
K479 & 21 29 56.79 & +12 10 27.0 & 13.99 & 12.68 & 10.276 & 9.678 & 9.524 & 1.31 & 3.16 & 4,5 \\
K482 & 21 29 56.94 & +12 08 44.7 & 15.25 & 14.38 & 12.632 & 12.195 & 12.071 & 0.87 & 2.31 & 3 \\ 
K506 & 21 29 57.43 & +12 08 21.5 & 15.62 & 14.75 & 12.848 & 12.299 & 12.232 & 0.87 & 2.52 & 4 \\
K550 & 21 29 58.03 & +12 11 54.2 & 16.00 & 15.10 & 13.357 & 12.833 & 12.742 & 0.90 & 2.36 & 2,3,4 \\
K567 & 21 29 58.33 & +12 09 12.8 & 14.46 & 13.25 & 11.129 & 10.508 & 10.414 & 1.21 & 2.84 & 3,5 \\
K582 & 21 29 58.60 & +12 08 08.0 & 15.32 & 14.48 & 12.829 & 12.370 & 12.305 & 0.84 & 2.18 & 1 \\
K583 & 21 29 58.57 & +12 09 21.4 & 14.26 & 12.83 & 10.315 & 9.726 & 9.569 & 1.43 & 3.26 & 1 \\
K647 & 21 29 59.46 & +12 08 35.6 & 14.79 & 13.60 & 11.388 & 10.807 & 10.686 & 1.19 & 2.91 & 4 \\
K654 & 21 29 59.53 & +12 11 52.6 & 16.06 & 15.16 & 13.350 & 12.861 & 12.785 & 0.90 & 2.38 & 4 \\
K672 & 21 29 59.81 & +12 11 10.7 & 14.84 & 13.84 & 11.837 & 11.290 & 11.211 & 1.00 & 2.63 & 3,5 \\
K677 & 21 29 59.99 & +12 06 26.6 & 15.82 & 15.00 & 13.330 & 12.824 & 12.811 & 0.82 & 2.19 & 3,4 \\
K691 & 21 30 00.03 & +12 13 39.5 & 15.64 & 14.80 & 12.859 & 12.321 & 12.217 & 0.84 & 2.58 & 1,2 \\
K702 & 21 30 00.34 & +12 10 50.9 & 14.47 & 12.99 & 10.609 & 10.080 & 9.910 & 1.48 & 3.08 & 1,2 \\
K709 & 21 30 00.38 & +12 07 36.4 & 14.75 & 13.61 & 11.535 & 10.944 & 10.847 & 1.14 & 2.76 & 3,4 \\
K736 & 21 30 00.63 & +12 09 28.4 & 15.02 & 14.01 & 12.063 & 11.474 & 11.346 & 1.01 & 2.66 & 3,5 \\
K757 & 21 30 00.91 & +12 08 57.1 & 14.31 & 12.88 & 10.383 & 9.759 & 9.605 & 1.43 & 3.28 & 1,2 \\
K800 & 21 30 01.65 & +12 12 30.3 & 16.23 & 15.12 & 13.029 & 12.575 & 12.761 & 1.11 & 2.36 & 1 \\
K825 & 21 30 02.25 & +12 11 21.5 & 14.13 & 12.79 & 10.227 & 9.582 & 9.433 & 1.34 & 3.36 & 3 \\
K846 & 21 30 02.78 & +12 06 55.7 & 15.05 & 13.97 & 11.918 & 11.388 & 11.247 & 1.08 & 2.72 & 1,4,5 \\
K853 & 21 30 02.74 & +12 10 43.9 & 14.27 & 12.88 & 10.469 & 9.860 & 9.727 & 1.39 & 3.15 & 1,2 \\
K866 & 21 30 03.09 & +12 10 21.8 & 15.72 & 14.70 & 12.885 & 12.322 & 12.269 & 1.02 & 2.43 & 3 \\
K875 & 21 30 03.17 & +12 13 28.7 & 15.05 & 14.10 & 12.180 & 11.659 & 11.506 & 0.95 & 2.60 & 1,2 \\
K879 & 21 30 03.50 & +12 03 12.5 & 15.22 & 14.16 & 12.205 & 11.615 & 11.527 & 1.06 & 2.63 & 1,2 \\
K902 & 21 30 04.00 & +12 08 57.8 & 15.90 & 14.84 & 13.003 & 12.483 & 12.364 & 1.06 & 2.48 & 3,4 \\
K906 & 21 30 04.09 & +12 07 27.1 & 16.24 & 15.37 & 13.560 & 13.046 & 13.009 & 0.87 & 2.36 & 3 \\
K919 & 21 30 04.32 & +12 10 56.2 & 14.80 & 13.60 & 11.459 & 10.893 & 10.757 & 1.20 & 2.84 & 1,2 \\
K925 & 21 30 04.62 & +12 08 53.7 & 15.61 & 14.62 & 12.733 & 12.231 & 12.067 & 0.99 & 2.55 & 1,2 \\
K926 & 21 30 04.65 & +12 07 40.5 & 15.89 & 15.18 & 13.591 & 13.186 & 13.107 & 0.71 & 2.07 & 4 \\
K932 & 21 30 04.75 & +12 11 10.3 & 15.14 & 14.09 & 12.099 & 11.534 & 11.438 & 1.05 & 2.65 & 4 \\
K947 & 21 30 05.18 & +12 13 20.3 & 15.26 & 14.29 & 12.311 & 11.724 & 11.613 & 0.97 & 2.68 & 2 \\
K954 & 21 30 05.54 & +12 08 55.3 & 15.32 & 14.33 & 12.357 & 11.835 & 11.719 & 0.99 & 2.61 & 4 \\
K969 & 21 30 06.37 & +12 06 59.3 & 14.61 & 13.45 & 11.364 & 10.829 & 10.700 & 1.16 & 2.75 & 1,4 \\
K979 & 21 30 06.96 & +12 07 46.5 & 15.09 & 14.23 & 12.454 & 11.978 & 11.898 & 0.86 & 2.33 & 1 \\
K989 & 21 30 07.30 & +12 10 50.7 & 16.02 & 15.13 & 13.556 & 13.078 & 12.950 & 0.89 & 2.18 & 3 \\
K993 & 21 30 07.40 & +12 10 33.1 & 15.01 & 14.01 & 11.964 & 11.399 & 11.264 & 1.00 & 2.75 & 1 \\
K1010 & 21 30 08.42 & +12 09 42.1 & 15.88 & 14.97 & 13.167 & 12.630 & 12.545 & 0.91 & 2.43 & 4 \\
K1014 & 21 30 08.95 & +12 08 49.1 & 15.61 & 14.70 & 13.003 & 12.483 & 12.364 & 0.91 & 2.34 & 3 \\
K1029 & 21 30 09.71 & +12 13 42.4 & 14.74 & 13.56 & 11.349 & 10.739 & 10.601 & 1.18 & 2.96 & 1,2 \\
K1030 & 21 30 09.78 & +12 12 54.4 & 15.17 & 14.14 & 12.109 & 11.538 & 11.401 & 1.03 & 2.74 & 1,2 \\
K1033 & 21 30 09.89 & +12 10 52.5 & 15.41 & 14.44 & 12.447 & 11.882 & 11.773 & 0.97 & 2.67 & 1,2 \\
K1040 & 21 30 10.49 & +12 10 06.2 & 14.60 & 13.40 & 11.151 & 10.562 & 10.438 & 1.20 & 2.96 & 3 \\
K1049 & 21 30 11.31 & +12 01 48.5 & 15.53 & 14.56 & 12.604 & 12.038 & 11.934 & 0.97 & 2.63 & 1 \\
K1054 & 21 30 11.38 & +12 08 41.2 & 15.19 & 14.20 & 12.181 & 11.626 & 11.517 & 0.99 & 2.68 & 1,2 \\
K1056 & 21 30 11.69 & +12 10 33.7 & 15.71 & 14.79 & 12.907 & 12.384 & 12.227 & 0.92 & 2.56 & 4 \\
K1069 & 21 30 14.26 & +12 09 23.4 & 15.42 & 14.60 & 12.660 & 12.125 & 12.006 & 0.82 & 2.60 & 4 \\
K1073 & 21 30 14.95 & +12 10 20.7 & 15.19 & 14.18 & 12.127 & 11.593 & 11.465 & 1.01 & 2.72 & 1,2,3,4 \\
K1074 & 21 30 15.23 & +12 11 34.5 & 16.22 & 15.26 & 13.406 & 12.910 & 12.786 & 0.96 & 2.47 & 1,3,4 \\
K1079 & 21 30 15.66 & +12 08 22.9 & 15.09 & 14.09 & 12.061 & 11.517 & 11.408 & 1.00 & 2.68 & 1,2 \\
K1083 & 21 30 15.78 & +12 16 59.6 & 16.19 & 15.43 & 13.566 & 13.053 & 12.987 & 0.76 & 2.44 & 1,2,3 \\
K1084 & 21 30 16.06 & +12 13 34.3 & 15.39 & 14.42 & 12.415 & 11.860 & 11.772 & 0.97 & 2.65 & 1,2,3 \\
K1097 & 21 30 21.04 & +12 13 00.8 & 16.21 & 15.41 & 13.631 & 13.127 & 13.025 & 0.80 & 2.39 & 1,2,3,4 \\
K1106 & 21 30 22.71 & +12 17 59.6 & 15.54 & 14.71 & 12.719 & 12.163 & 12.044 & 0.83 & 2.67 & 2,3 \\
K1136 & 21 30 31.78 & +12 08 54.8 & 15.70 & 14.87 & 13.005 & 12.483 & 12.329 & 0.83 & 2.54 & 1,2,3,4 \\
\enddata
\tablenotetext{a}{\citet{kustner01} is the identification for the majority of the stars denoted by K.}
\tablenotetext{b}{2MASS coordinates \citep{skrutskie01}.}
\tablenotetext{c}{Observations: 1: 2005 May 22; 2: 2006 May 11; 3: 2006 October 4; 4: 2006 October 7; 5: Dusty giants
identified by \citet{boyer01}.}
\tablenotetext{d}{\citet{brown01}.}
\tablenotetext{e}{\citet{cudworth01}.}
\tablenotetext{f}{\citet{gebhardt01}.}
\tablenotetext{g}{B and V magnitudes are taken from \citet{auriere01}.}
\tablecomments{The visual photometry is taken from \citet{cudworth01}; 
J,H,K photometry is taken from the 2MASS Catalog \citep{skrutskie01}.}
\end{deluxetable*}

\clearpage

\begin{deluxetable*}{lcccc}
\tabletypesize{\scriptsize}
\tablecaption{Hectochelle Observations of M15}
\tablewidth{0pt}
\tablehead{
\colhead{Date}           & \colhead{Total exp.}      &
\colhead{Wavelength} & \colhead{Filter Name}  & \colhead{Number of}   \\
\colhead{(UT)} & \colhead{(s)} & \colhead{(\AA)} & \colhead{} & \colhead{Observed Stars} 
}
\startdata
2005 May 22 (Field 1)& $3 \times 1200$ & 6485$-$6575 & OB25 & 53 \\
2005 May 23 (Field 1)& $3 \times 1200$ & 3910$-$3990 & Ca41 & 53 \\
2006 May 11 (Field 2)& $3 \times 2100$ & 6475$-$6630 & OB25 & 54 \\
2006 October 4 (Field 3) & $3 \times 2100$ & 6475$-$6630 & OB25 & 58 \\
2006 October 7 (Field 4) & $3 \times 2100$ & 6475$-$6630 & OB25 & 50 \\
\enddata
\end{deluxetable*}

\clearpage

\begin{deluxetable*}{lccccccc}
\tablecolumns{7}
\tabletypesize{\scriptsize}
\tablecaption{Physical Parameters of Cluster Members}
\tablewidth{0pt}
\tablehead{
\colhead{ID No.}           & \colhead{$M_{V}$}      &
\colhead{$(B-V)_0$}          & \colhead{$(V-K)_0$}  &
\colhead{P \tablenotemark{a}}          & \colhead{$T_{eff}$} & \colhead{$log L/L_{\odot}$} & \colhead{$R/R_{\odot}$} \\
\colhead{} & \colhead{} & \colhead{} & \colhead{} & \colhead{} & \colhead{(K)} & \colhead{}  & \colhead{}
}
\startdata
B5 & $-1.86$ & 1.02 & 2.595 & 99  & 4490 & 2.850 	   & 42.9	   \\
B6 & $-1.83$ & 1.05 & 2.605 & 98  & 4480 & 2.840 	   & 42.6	   \\
B16 & $-1.61$ & 0.99 & 2.455 & 0  & 4610 & 2.724 	   & 35.2	   \\
B30 & $-1.68$ & 1.01 & 2.605 & 99  & 4480 & 2.780 	   & 39.7	   \\
C3 & $-0.72$ & 0.74 & 2.335 & 97  & 4730 & 2.348 	   &   21.7     \\
C20 & +0.09 & 0.79 & 2.195 & 97  & 4870 & 2.004 	   & 13.8	   \\
C35 & $-0.23$ & 0.53 & 2.015 & 0  & 5090 & 2.109 	   & 14.2	   \\
GEB 254 & $-2.13$ & 1.21 & 3.195 & \nodata  & 4080 & 3.052 & 65.5	   \\
GEB 289 & $-1.87$ & 0.94 & 2.845 & \nodata & 4300 & 2.864  &  47.5	   \\
K12 & $-0.32$ & 0.69 & 1.925 & 99  & 5200 & 2.135 	   & 14.0	   \\
K21 & $-0.37$ & 0.64 & 1.995 & 72  & 5110 & 2.163 	   &  15.0	     \\
K22 & $-0.96$ & 0.91 & 2.325 & 96  & 4740 & 2.443 	   &  24.1	     \\
K26 & $-0.37$ & 0.77 & 2.155 & 92  & 4920 & 2.182 	   &  16.5	     \\
K27 & $-0.28$ & 0.65 & 1.925 & 88  & 5200 & 2.119 	   &  13.8	     \\
K31 & $-0.16$ & 0.73 & 2.055 & 76  & 5040 & 2.086 	   &  14.1	     \\
K42 & $-0.14$ & 0.77 & 2.065 & 85  & 5020 & 2.079 	   &  14.1	     \\
K47 & $-1.26$ & 0.92 & 2.345 & 99  & 4720 & 2.566 	   &  28.0	     \\
K56 & +0.00 & 0.80 & 2.125 & 93  & 4950 & 2.030 	   &  13.7	  \\
K60 & $-0.11$ & 0.71 & 2.055 & 86  & 5040 & 2.066 	   &  13.8	     \\
K64 & $-0.12$ & 0.76 & 2.095 & 75  & 4990 & 2.074 	   &  14.2	  \\
K69 & $-0.92$ & 0.81 & 2.325 & 99  & 4740 & 2.427 	   &  23.6	  \\
K70 & $-1.05$ & 0.85 & 2.395 & 3  & 4670 & 2.490 	   &  26.2	     \\
K77 & $-1.55$ & 1.00 & 2.515 & 99  & 4560 & 2.711 	   &  35.4	  \\
K87 & $-1.57$ & 0.97 & 2.455 & 99  & 4610 & 2.708 	   &  34.5	  \\
K89 & $-0.84$ & 0.72 & 2.305 & 99  & 4760 & 2.392 	   &  22.5	  \\
K92 & $-0.17$ & 0.84 & 2.135 & 95  & 4940 & 2.100 	   &  14.9	  \\
K105 & $-0.08$ & 0.87 & 2.075 & 89  & 5010 & 2.056 	   & 13.8	   \\
K112 & $-0.29$ & 0.78 & 2.215 & 94  & 4850 & 2.158 	   & 16.6	   \\
K114 & $-1.52$ & 1.04 & 2.495 & 99  & 4580 & 2.695 	   & 34.5	   \\
K129 & $-1.11$ & 0.87 & 2.365 & 99  & 4700 & 2.509 	   & 26.4	\\
K133 & $-0.27$ & 0.79 & 1.925 & 82  & 5200 & 2.115 	   & 13.7	\\
K136 & $-0.37$ & 0.70 & 2.015 & 53  & 5090 & 2.165 	   & 15.2	\\
K137 & +0.01 & 0.88 & 2.045 & 93  & 5050 & 2.016 	   & 13.0	   \\
K144 & $-2.31$ & 1.24 & 2.845 & 99  & 4300 & 3.083 	   & 61.1	   \\
K145 & $-0.01$ & 0.78 & 2.125 & 83  & 4950 & 2.034 	   & 13.8	   \\
K146 & $-1.80$ & 1.02 & 2.565 & 99  & 4520 & 2.820 	   & 40.9	   \\
K151 & $-0.26$ & 0.79 & 2.335 & 92  & 4730 & 2.164 	   & 17.5	\\
K152 & $-0.12$ & 0.76 & 2.095 & 76  & 4990 & 2.074 	   & 14.2	\\
K153 & +0.11 & 0.65 & 2.135 & 62  & 4940 & 1.988 	   & 13.1	\\
K158 & $-1.21$ & 0.78 & 2.075 & 99  & 5010 & 2.508 	   & 23.2	\\
K202 & $-0.20$ & 0.79 & 2.055 & 64  & 5040 & 2.102 	   & 14.4	   \\
K224 & $-1.98$ & 1.10 & 2.605 & 99  & 4480 & 2.900 	   & 45.6	   \\
K238 & $-2.13$ & 1.18 & 2.665 & 99  & 4390 & 2.972 	   &  51.6	    \\
K255 & $-1.61$ & 0.99 & 2.455 & 99  & 4610 & 2.724 	   &  35.2	    \\
K260 & $-1.38$ & 0.70 & 1.665 & 99  & 5590 & 2.536 	   &  19.3	    \\
K272 & $-1.89$ & 1.09 & 2.595 & 99  & 4460 & 2.862 	   &  44.1   	 \\
K288 & $-1.66$ & 1.04 & 2.465 & 99  & 4600 & 2.746 	   &  36.2   	 \\
K328 & $-1.67$ & 1.06 & 2.525 & 99  & 4550 & 2.760 	   & 37.6	 \\
K337 & $-0.61$ & 0.89 & 2.285 & 92  & 4780 & 2.297 	   &  20.0	    \\
K341 & $-2.56$ & 1.27 & 2.845 & 99  & 4300 & 3.183 	   &  68.6	    \\
K361 & $-1.41$ & 0.91 & 2.325 & 99  & 4740 & 2.623 	   &  29.6	    \\
K393 & $-2.12$ & 1.11 & 2.665 & 99  & 4430 & 2.967 	   &  50.4	    \\
K421 & $-2.65$ & 1.10 & 2.795 & 99  & 4330 & 3.207 	   &  69.5	    \\
K431 & $-2.34$ & 1.20 & 2.715 & 99  & 4390 & 3.066 	   &  57.5    \\
K447 & $-2.12$ & 0.97 & 2.505 & 99  & 4570 & 2.937 	   &  45.7	    \\
K462 & $-2.47$ & 1.45 & 2.905 & 99  & 4260 & 3.161 	   &  68.1	    \\
K476 & $-0.74$ & 0.82 & 2.225 & 98  & 4840 & 2.340 	   &  20.5	 \\
K479 & $-2.69$ & 1.21 & 2.885 & 99  & 4270 & 3.244 	   &  74.6	 \\
K482 & $-0.99$ & 0.77 & 2.035 & 99  & 5060 & 2.415 	   &  20.5	    \\
K506 & $-0.62$ & 0.77 & 2.245 & 99  & 4820 & 2.295 	   &  19.6	 \\
K550 & $-0.27$ & 0.80 & 2.085 & 93  & 5000 & 2.133 	   &  15.1	 \\
K567 & $-2.32$ & 1.11 & 2.565 & 99  & 4520 & 3.028 	   &  51.9	 \\
K582 & $-0.89$ & 0.74 & 1.905 & 99  & 5230 & 2.361 	   &  18.0  	 \\
K583 & $-2.54$ & 1.33 & 2.985 & 99  & 4200 & 3.210 	   &  74.2	    \\
K647 & $-1.77$ & 1.09 & 2.635 & 99  & 4460 & 2.821 	   &  42.0  	\\
K654 & $-0.21$ & 0.80 & 2.105 & 89  & 4980 & 2.112 	   &  14.9  	\\
K672 & $-1.53$ & 0.90 & 2.355 & 99  & 4710 & 2.676 	   &  31.9  	\\
K677 & $-0.37$ & 0.72 & 1.915 & 98  & 5220 & 2.154 	   &  14.2  	\\
K691 & $-0.57$ & 0.74 & 2.305 & 99  & 4760 & 2.284 	   &  19.9	    \\
K702 & $-2.38$ & 1.38 & 2.805 & 99  & 4330 & 3.102 	   &  61.6   	 \\
K709 & $-1.76$ & 1.04 & 2.485 & 99  & 4590 & 2.789 	   &  38.2   	 \\
K736 & $-1.36$ & 0.91 & 2.385 & 99  & 4680 & 2.613 	   &  30.0	 \\
K757 & $-2.49$ & 1.33 & 3.005 & 99  & 4190 & 3.195 	   &  73.2	 \\
K800 & $-0.25$ & 1.01 & 2.085 & 65  & 5000 & 2.125 	   &  15.0	 \\
K825 & $-2.58$ & 1.24 & 3.085 & 99  & 4140 & 3.253 	   &  80.2	    \\
K846 & $-1.40$ & 0.98 & 2.445 & 99  & 4620 & 2.638 	   &  31.7	 \\
K853 & $-2.49$ & 1.29 & 2.875 & 99  & 4280 & 3.162 	   &  67.6	    \\
K866 & $-0.67$ & 0.92 & 2.155 & 86  & 4920 & 2.302 	   &  19.0	    \\
K875 & $-1.27$ & 0.85 & 2.325 & 99  & 4740 & 2.567 	   &  27.8	    \\
K879 & $-1.21$ & 0.96 & 2.355 & 98  & 4710 & 2.547 	   &  27.5	    \\
K902 & $-0.53$ & 0.96 & 2.205 & 97  & 4860 & 2.253 	   &  18.4	    \\
K906 & +0.00 & 0.77 & 2.085 & 88  & 5000 & 2.025 	   &  13.4	    \\  	    
K919 & $-1.77$ & 1.10 & 2.565 & 99  & 4520 & 2.808 	   &  40.3	 \\
K925 & $-0.75$ & 0.89 & 2.275 & 96  & 4790 & 2.351 	   &  21.2	 \\
K926 & $-0.19$ & 0.61 & 1.795 & 84  & 5390 & 2.070 	   &  12.1	 \\
K932 & $-1.28$ & 0.95 & 2.375 & 99  & 4690 & 2.579 	   &  28.8	    \\
K947 & $-1.08$ & 0.87 & 2.405 & 99  & 4660 & 2.504 	   &  26.7  	\\
K954 & $-1.04$ & 0.89 & 2.335 & 99  & 4730 & 2.476 	   &  25.1  	\\
K969 & $-1.92$ & 1.06 & 2.475 & 99  & 4590 & 2.851 	   &  41.1	    \\
K979 & $-1.14$ & 0.76 & 2.055 & 99  & 5040 & 2.478 	   &  22.2	    \\
K989 & $-0.24$ & 0.79 & 1.905 & 73  & 5230 & 2.101 	   &  13.3  	\\
K993 & $-1.36$ & 0.90 & 2.475 & 99  & 4590 & 2.627 	   &  31.7  	\\
K1010 & $-0.40$ & 0.81 & 2.155 & 97  & 4920 & 2.194 	   &  16.8	    \\
K1014 & $-0.67$ & 0.81 & 2.065 & 98  & 5020 & 2.291 	   & 18.0	    \\
K1029 & $-1.81$ & 1.08 & 2.685 & 99  & 4420 & 2.848 	   &  44.1	    \\
K1030 & $-1.23$ & 0.93 & 2.465 & 99  & 4600 & 2.574 	   & 29.7	    \\
K1033 & $-0.93$ & 0.87 & 2.395 & 99  & 4670 & 2.442 	   &  24.8    \\
K1040 & $-1.97$ & 1.10 & 2.685 & 99  & 4420 & 2.912 	   &  47.5	    \\
K1049 & $-0.81$ & 0.87 & 2.355 & 0  & 4710 & 2.387 	   &  22.9	    \\
K1054 & $-1.17$ & 0.89 & 2.405 & 99  & 4660 & 2.540	   &  27.9	    \\
K1056 & $-0.58$ & 0.82 & 2.285 & 94  & 4780 & 2.285 	   &  19.7	 \\
K1069 & $-0.77$ & 0.72 & 2.325 & 99  & 4740 & 2.367 	   &  22.1	    \\
K1073 & $-1.19$ & 0.91 & 2.445 & 99  & 4620 & 2.554 	   &  28.8	    \\
K1074 & $-0.11$ & 0.86 & 2.195 & 88  & 4870 & 2.084 	   &  15.1	    \\
K1079 & $-1.28$ & 0.90 & 2.405 & 99  & 4660 & 2.584 	   &  29.3   	 \\
K1083 & +0.06 & 0.66 & 2.165 & 66  & 4910 & 2.012 	   &  13.7	    \\
K1084 & $-0.95$ & 0.87 & 2.375 & 99  & 4690 & 2.447 	   &  24.7	    \\
K1097 & +0.04 & 0.70 & 2.115 & 50  & 4970 & 2.013 	   &  13.3	    \\
K1106 & $-0.66$ & 0.73 & 2.395 & 96  & 4670 & 2.334 	   &  21.9	    \\
K1136 & $-0.50$ & 0.73 & 2.265 & 79  & 4800 & 2.250 	   &  18.8	    \\
\enddata						   
\tablenotetext{a}{Membership probability from proper motion observations \citep{cudworth01}.}
\end{deluxetable*}					   						   
		
\clearpage		
							   
\begin{deluxetable*}{lcccc}				   
\tabletypesize{\scriptsize}				   
\tablecaption{B/R ratio of H$\alpha$ Line for Stars with Emission Wings}
\tablewidth{0pt}					   
\tablehead{						   
\colhead{ID No.}           & \colhead{$B/R$ 2005 May 22}      &
\colhead{$B/R$ 2006 May 11} & \colhead{$B/R$ 2006 October 4} & \colhead{$B/R$ 2006 October 7}
}							   
\startdata
B5 & no emission & \nodata & no emission & $>1$ \\
B6 & no emission & $>1$ & no emission & no emission \\
B30 & no emission & no emission & $>1$ & $>1$ \\
GEB 254 & \nodata & \nodata & \nodata & $>1$ \\
K144 & \nodata & $>1$ & no emission & $>1$ \\
K158 & \nodata & $>1$ & \nodata & \nodata \\
K224 & $>1$ & $>1$ & $>1$ & \nodata \\
K238 & \nodata & \nodata & \nodata & $>1$ \\
K260 & $<1$ & $>1$ & \nodata & \nodata \\
K341 & $<1$ & $>1$ & $>1$ & $>1$ \\
K393 & $>1$ & \nodata & \nodata & \nodata \\
K421 & \nodata & \nodata & $>1$ & \nodata \\
K431 & no emission & $>1$ & $>1$ & \nodata \\
K447 & \nodata & $<1$ & \nodata & \nodata \\
K462 & \nodata & $<1$ & \nodata & \nodata \\
K479 & \nodata & \nodata & \nodata & $<1$\\
K482 & \nodata & \nodata & $>1$ & \nodata \\
K582 & $<1$ & \nodata & \nodata & \nodata \\
K672 & \nodata & \nodata & $>1$ & \nodata \\
K702 & no emission & $>1$ & \nodata & \nodata \\
K709 & \nodata & \nodata & \nodata & $>1$ \\
K757 & $>1$ & $>1$ & \nodata & \nodata \\
K825 & \nodata & \nodata & $<1$ & \nodata \\
K853 & $>1$ & no emission & \nodata & \nodata \\
K875 & $>1$ & no emission & \nodata & \nodata \\
K969 & $>1$ & \nodata & \nodata & $<1$ \\
K979 & $<1$ & \nodata & \nodata & \nodata \\
K1029 & $>1$ & no emission & \nodata & \nodata \\
K1040 & \nodata & \nodata & $>1$ & \nodata \\
\enddata
\tablecomments{The parameter B/R is the intensity 
ratio of Blue (short wavelength) and Red (long wavelength) emission peaks. The symbol \nodata indicates the star was not
observed. If B/R ratio is $>1$ the line profile indicates inflow, if B/R ratio is $<1$ the line profile indicates outflow.}
\end{deluxetable*}

\clearpage

\begin{deluxetable*}{lcccc}
\tabletypesize{\scriptsize}
\tablecaption{Radial Velocity of Cluster Members}
\tablewidth{0pt}
\tablehead{
\colhead{ID No.} &
\colhead{$v_{rad, 1}$ \tablenotemark{a}}         & 
\colhead{$v_{rad, 2}$ \tablenotemark{a}}        & 
\colhead{$v_{rad, 3}$ \tablenotemark{a}}  &
\colhead{$v_{rad, 4}$ \tablenotemark{a}}  \\
\colhead{} & \colhead{\tiny{(km \ s$^{-1}$)}} & \colhead{\tiny{(km \ s$^{-1}$)}} & \colhead{\tiny{(km \ s$^{-1}$)}} &
\colhead{\tiny{(km \ s$^{-1}$)}} 
}
\startdata
B5 &  $-109.1 \ \pm$ 0.3 & \nodata &  	$-102.3 \ \pm$ 0.4 & $-108.4 \ \pm$  0.3\\
B6 &  $-113.8 \ \pm$ 0.3 & $-113.3 \ \pm$ 0.3 & $-112.4 \ \pm$ 0.3& $-113.3 \ \pm$  0.4  \\
B16 & \nodata & \nodata & 		$-103.4 \ \pm$ 0.4 & \nodata \\
B30 & $-104.2 \ \pm$ 0.3 & $-105.7 \ \pm$ 0.3 & $-102.8 \ \pm$ 0.4& $-104.2 \ \pm$  0.3\\
C3 &  $-102.3 \ \pm$ 0.5 & $-101.3 \ \pm$ 0.4 & 	$-101.2 \ \pm$ 0.5& $-101.1 \ \pm$  0.5\\
C20 & $-108.0 \ \pm$ 0.6 & $-109.6 \ \pm$ 0.7 & $-108.6 \ \pm$ 0.6& $-108.6 \ \pm$  0.7\\
C35 & $-106.8 \ \pm$ 0.6 & \nodata &  \nodata 	   & \nodata \\
GEB 254 & \nodata & \nodata & \nodata & 		$-100.7  \ \pm$ 0.3 \\
GEB 289 & \nodata & \nodata &		 $-100.4 \ \pm$  0.5 & \nodata \\
K12 & $-107.6 \ \pm$ 0.8 & $-107.0 \ \pm$ 0.6 & $-107.2 \ \pm$ 0.8& $-107.0 \ \pm$  0.9\\
K21 & \nodata & $-112.0 \ \pm$ 0.7 &  	$-111.0 \ \pm$ 0.8& $-112.9 \ \pm$  0.7\\
K22 & $-101.7 \ \pm$ 0.4 & $-101.2 \ \pm$ 0.3 &  $-102.1 \ \pm$ 0.4& $-102.0 \ \pm$  0.4  \\
K26 & $-104.0 \ \pm$ 0.5 & $-104.0 \ \pm$ 0.4 & $-104.0 \ \pm$ 0.5 & $-103.7 \ \pm$  0.5\\
K27 & \nodata & $-107.4 \ \pm$ 0.6 & 	$-106.9 \ \pm$ 0.7 & $-107.2 \ \pm$  0.8\\
K31 & \nodata & $-105.3 \ \pm$ 0.4 & 	$-105.5 \ \pm$ 0.5& $-105.8 \ \pm$  0.6\\
K42 & \nodata &\nodata & 		$-104.5 \ \pm$ 0.7 & $-104.3 \ \pm$  0.8\\
K47 & $-102.9 \ \pm$ 0.7 & $-102.0 \ \pm$ 0.4 & 	\nodata 	   & \nodata \\
K56 & $-104.7 \ \pm$ 0.7 & $-104.4 \ \pm$ 0.5 & $-105.4 \ \pm$ 0.7& $-104.3 \ \pm$  0.7\\
K60 & \nodata & \nodata & 		$-109.5 \ \pm$ 0.5 & $-109.2 \ \pm$  0.6\\
K64 & $-107.8 \ \pm$ 0.8 & $-108.6 \ \pm$ 0.6 & $-108.8 \ \pm$ 0.7& $-108.9 \ \pm$  0.7\\
K69 & $-102.7 \ \pm$ 0.6 & $-103.1 \ \pm$ 0.5 &  \nodata 		   & $-103.1 \ \pm$  0.6\\
K70 & $-109.2 \ \pm$ 0.4 & \nodata &  		\nodata 	 & \nodata \\
K77 & $-104.7 \ \pm$ 0.4 & $-103.7 \ \pm$ 0.3 & $-104.2 \ \pm$ 0.4& $-104.3 \ \pm$  0.4\\
K87 & \nodata & \nodata & \nodata 	&		$-108.3 \ \pm$  0.4 \\
K89 & \nodata & $-109.5 \ \pm$ 0.4 &  	$-109.8 \ \pm$ 0.5& $-109.8 \ \pm$  0.6\\
K92 & $-107.5 \ \pm$ 0.9 & $-106.7 \ \pm$ 0.3 & $-106.5 \ \pm$ 0.5 & $-106.1 \ \pm$  0.6 \\
K105 & $-112.3 \ \pm$ 0.5 & \nodata & 	\nodata 		 & $-112.2  \ \pm$ 0.6\\
K112 & \nodata & \nodata & \nodata 	&		$-100.3 \ \pm$  0.5 \\
K114 & \nodata & \nodata & 		$-113.2 \ \pm$ 0.4 & \nodata \\
K129 & \nodata & \nodata & 		$-102.5 \ \pm$ 0.4 & \nodata  \\
K133 & \nodata & $-104.3 \ \pm$ 0.5 & 	$-104.6 \ \pm$ 0.7 & \nodata \\
K136 & \nodata & \nodata & \nodata 	&		$-111.0  \ \pm$ 0.7 \\
K137 & \nodata & \nodata & 		$-111.5 \ \pm$ 0.6 & \nodata \\
K144 & \nodata & $-108.6 \ \pm$ 0.3 & 	 	\nodata 	   & $-110.9 \ \pm$  0.3 \\
K145 & $-110.7 \ \pm$ 0.6 & $-110.7 \ \pm$ 0.4 &$-111.1 \ \pm$ 0.6  & \nodata \\
K146 & \nodata & $-101.3 \ \pm$ 0.3 & 	 \nodata 		 & \nodata \\
K151 & \nodata & \nodata & 		$-94.1  \ \pm$0.5 & \nodata \\
K152 & $-100.8 \ \pm$ 0.6 & $-100.0 \ \pm$ 0.4 & $-100.7 \ \pm$ 0.4 & \nodata \\
K153 & $-110.3 \ \pm$ 0.6 & \nodata & 	$-109.9 \ \pm$ 0.8& $-111.0 \ \pm$  0.7 \\
K158 & \nodata & $-110.5 \ \pm$ 0.5 & 		\nodata 	   & \nodata \\
K202 & \nodata & \nodata & 		 $-99.4 \ \pm$ 0.7 & \nodata \\
K224 & $-106.6 \ \pm$ 0.3 & $-106.7 \ \pm$ 0.3 & $-106.8 \ \pm$  0.4& \nodata \\
K238 & \nodata & \nodata & \nodata & 			$-102.0  \ \pm$ 0.3 \\
K255 & $-102.4 \ \pm$ 0.3 & $-101.9 \ \pm$ 0.4 & \nodata 		       & \nodata \\
K260 & $-96.5 \ \pm$ 1.4 & $-98.7 \ \pm$ 0.9 &  	\nodata 	       & \nodata \\
K272 & \nodata & \nodata & 		 $-106.2 \ \pm$  0.4 & \nodata \\
K288 & $-105.2 \ \pm$ 0.4 & \nodata & 	 $-104.4 \ \pm$  0.4& \nodata \\
K328 & \nodata & $-102.9 \ \pm$ 0.3 & 	\nodata 	 	       & $-102.4  \ \pm$ 0.4\\
K337 & \nodata & $-107.2 \ \pm$ 0.4 & 	 $-107.8 \ \pm$ 0.6& $-107.8  \ \pm$ 0.6 \\
K341 & $-111.8 \ \pm$ 0.2 & $-111.3 \ \pm$ 0.3 & $-110.9 \ \pm$ 0.3 & $-110.9  \ \pm$ 0.3 \\
K361 & \nodata & \nodata & 		 $-108.6 \ \pm$ 0.5 & \nodata \\
K393 & $-96.4 \ \pm$ 0.3 & \nodata & 	\nodata 	 	      & \nodata \\
K421 & \nodata & \nodata & 		 $-111.7 \ \pm$ 0.3 & \nodata \\
K431 & $-107.0 \ \pm$ 0.3 & $-105.8 \ \pm$ 0.3 & $-107.1 \ \pm$ 0.4 & \nodata \\
K447 & \nodata & $-105.2 \ \pm$ 0.3 & 	\nodata 		& \nodata \\
K462 & \nodata & $-113.4 \ \pm$ 0.3 & 	\nodata 		 & \nodata \\
K476 & \nodata & $-109.1 \ \pm$ 0.5 & 	\nodata 		& \nodata \\
K479 & \nodata & \nodata & \nodata & 		$-122.5  \ \pm$ 0.4 \\
K482 & \nodata & \nodata & 		 $-110.2 \ \pm$ 0.5 & \nodata \\
K506 & \nodata & \nodata & \nodata & 		$-103.5  \ \pm$ 0.5 \\
K550 & \nodata & $-111.1 \ \pm$ 0.5 & 	 $-110.2 \ \pm$ 0.7& $-110.2  \ \pm$ 0.7 \\
K567 & \nodata & \nodata & 		 $-93.42 \ \pm$ 0.4 & \nodata \\
K582 & $-99.4 \ \pm$ 0.7 & \nodata & 		\nodata     		& \nodata \\
K583 & $-109.2 \ \pm$ 0.3 & \nodata & 	\nodata 	 	   & \nodata \\
K647 & \nodata & \nodata & \nodata & 		$-116.8  \ \pm$ 0.4 \\
K654 & \nodata & \nodata & \nodata & 		$-109.5 \ \pm$  0.7 \\
K672 & \nodata & \nodata &		 $-106.9 \ \pm$ 0.5 & \nodata \\
K677 & \nodata & \nodata & 		 $-104.9 \ \pm$ 0.8 & $-105.2  \ \pm$ 0.9 \\
K691 & $-109.4 \ \pm$ 0.4 & $-109.6 \ \pm$ 0.4 & 	\nodata  	       & \nodata \\
K702 & $-116.7 \ \pm$ 0.2 & $-117.9 \ \pm $ 0.3 & 	\nodata 		& \nodata \\
K709 & \nodata & \nodata &		 $-101.7 \ \pm$ 0.4   & $-99.8  \ \pm$ 0.4\\
K736 & \nodata & \nodata & 		 $-99.7  \ \pm$ 0.4    & \nodata \\
K757 & $-117.4 \ \pm$ 0.3 & $-111.2 \ \pm $ 0.3 & \nodata 	     	& \nodata \\
K800 & $-104.4 \ \pm$ 0.8 & \nodata & 	\nodata 	 	    & \nodata \\
K825 & \nodata & \nodata & 		 $-101.4 \ \pm$ 0.4 & \nodata \\
K846 & $-105.5 \ \pm$ 0.3 & \nodata & 	\nodata  		      & $-104.6  \ \pm$ 0.4\\
K853 & $-108.3 \ \pm$ 0.3 & $-109.2 \ \pm$ 0.3 & \nodata 	       	& \nodata \\
K866 & \nodata & \nodata &		 $-109.6 \ \pm$ 0.7 & \nodata \\
K875 & $-111.2 \ \pm$ 0.4 & $-110.7 \ \pm$ 0.4 & \nodata 		 & \nodata \\
K879 & $-103.8 \ \pm$ 0.4 & $-104.2 \ \pm$ 0.3 & \nodata 		 & \nodata \\
K902 & \nodata & \nodata &		 $-108.4 \ \pm$  0.6 & $-109.0  \ \pm$ 0.7 \\
K906 & \nodata & \nodata & 		 $-106.3 \ \pm$  0.6 & \nodata \\
K919 & $-111.1 \ \pm$ 0.3 & $-112.4 \ \pm$ 0.3 &  \nodata 		& \nodata \\
K925 & $-108.4 \ \pm$ 0.5 & $-108.1 \ \pm$ 0.4 & \nodata 		 & \nodata \\
K926 & \nodata & \nodata & \nodata & 		$-108.1  \ \pm$ 1.0 \\
K932 & \nodata & \nodata & \nodata & 		$-107.8  \ \pm$ 0.5 \\
K947 & \nodata & $-116.8 \ \pm$ 0.4 & 	 \nodata 		& \nodata \\
K954 & \nodata & \nodata & \nodata & 		$-104.1  \ \pm$ 0.5 \\
K969 & $-110.8 \ \pm$ 0.4 & \nodata & 	\nodata  		& 	$-108.8  \ \pm$ 0.5 \\
K979 & $-111.2 \ \pm$ 0.6 & \nodata & 	\nodata  		&\nodata  \\
K989 & \nodata & \nodata & 		 $-109.3 \ \pm$  0.6 & \nodata \\
K993 & $-112.7 \ \pm$ 0.3 & \nodata &  	\nodata  		& \nodata \\
K1010 & \nodata & \nodata & \nodata & 		$-108.3 \ \pm$  0.8 \\
K1014 & \nodata & \nodata &		 $-115.4 \ \pm$  0.5 & \nodata \\
K1029 & $-101.6 \ \pm$ 0.4 & $-102.1 \ \pm$ 0.3 & \nodata 		& \nodata \\
K1030 & $-100.1 \ \pm$ 0.3 & $-101.1 \ \pm$ 0.3 & \nodata 		& \nodata \\
K1033 & $-111.4 \ \pm$ 0.4 & $-111.7 \ \pm$ 0.3 & \nodata 		& \nodata \\
K1040 & \nodata & \nodata &		 $-100.2 \ \pm$  0.3 & \nodata \\
K1049 & $-106.5 \ \pm$ 0.7 & \nodata & 	 \nodata 		& \nodata \\
K1054 & $-105.9 \ \pm$ 0.4 & $-106.5 \ \pm$ 0.4 &  \nodata 		& \nodata \\
K1056 & \nodata & \nodata & \nodata & 		$-103.5 \ \pm$  0.6 \\
K1069 & \nodata & \nodata & \nodata & 		$-102.4 \ \pm$  0.4 \\
K1073 & $-118.8 \ \pm$ 0.4 & $-120.3 \ \pm$ 0.4 & $-119.9 \ \pm$  0.5 & $-119.9 \ \pm$  0.5\\
K1074 & $-106.7 \ \pm$ 0.5 & \nodata &  	  $-106.9 \ \pm$  0.6& $-107.5  \ \pm$ 0.7\\
K1079 & $-104.4 \ \pm$ 0.4 & $-106.2 \ \pm$ 0.3 &  \nodata 		& \nodata \\
K1083 & $-107.1 \ \pm$ 0.6 & $-106.2 \ \pm$ 0.6 & $-106.5 \ \pm$  0.7 & \nodata \\
K1084 & $-106.3 \ \pm$ 0.4 & $-105.0 \ \pm$ 0.4 & $-103.7 \ \pm$  0.5 & \nodata \\
K1097 & $-109.9 \ \pm$ 0.5 & $-110.6 \ \pm$ 0.5 & $-108.5 \ \pm$  0.6& $-108.8  \ \pm$ 0.6\\
K1106 & \nodata & $-107.1 \ \pm$ 0.4 &  	 $-106.0 \ \pm$  0.5&\nodata  \\
K1136 & $-103.8 \ \pm$ 0.5 & $-105.3 \ \pm$ 0.4 & 	$-106.8 \ \pm$ 0.5 	& 	$-105.5 \ \pm$  0.6\\
\enddata
\tablenotetext{a}{Observations: 1: 2005 May 22, 2: 2006 May 11, 3: 2006 October 4, 4: 2006 October 7.}
\tablecomments{Data taken on 2006 May 11 and on 2006 October 7 were corrected with a velocity offset of +1.9~km~s$^{-1}$ and
+0.9~km~s$^{-1}$ respectively.}
\end{deluxetable*}

\clearpage

\begin{deluxetable*}{lccccrcc}
\tabletypesize{\scriptsize}
\tablecaption{Radial Velocities of Apparent Non-members}
\tablewidth{0pt}
\tablehead{
\colhead{ID No.}           & \colhead{$RA(2000)$ \tablenotemark{a} }    &
\colhead{Dec(2000) \tablenotemark{a} }          & \colhead{V \tablenotemark{b}} &
\colhead{B-V \tablenotemark{b}}          & \colhead{$v_{rad} $ \tiny{(km s$^{-1}$)} \tablenotemark{c}}      &
\colhead{P \tablenotemark{b}}  & \colhead{Obs. \tablenotemark{d} }
}
\startdata
B14 & 21 29 56.70 & +12 22 20.1 & 12.71 & 0.49 & $-22.7 \ \pm$ 0.4 & 99 & 1,2,4 \\
B22 & 21 30 36.04 & +12 05 17.6 & 13.92 & 0.91 & $-12.1 \ \pm$ 0.4 & 96 & 1,2,3,4 \\
B25 & 21 30 39.65 & +12 05 23.5 & 12.39 & 1.14 & $-5.4 \ \pm$ 0.4 & 99 & 1,2 \\
C19 & 21 29 52.30 & +11 59 40.2 & 14.89 & 0.87 & $-50.9 \ \pm$ 0.6 & 93 & 1,2,3,4 \\
K7 & 21 29 27.03 & +12 07 26.9 & 12.83 & 0.88 & $-177.4 \ \pm$ 0.5 & 98 & 1,2,4 \\
K28 & 21 29 35.27 & +12 14 40.0 & 13.67 & 1.04 & $-68.5 \ \pm$ 0.4 & 90 & 1,3,4 \\
K44 & 21 29 58.32 & +12 09 56.5 & 15.36 & 1.04 & $-41.4 \ \pm$ 0.5 & 67 & 3 \\
K73 & 21 29 44.19 & +12 09 17.1 & 13.62 & 0.74 & $-38.4 \ \pm$ 0.6 & 94 & 3 \\
K609 & 21 29 58.84 & +12 17 29.4 & 14.88 & 0.79 & $+4.7 \ \pm$ 0.6 & 96 & 1,2,3,4 \\
K996 & 21 30 06.80 & +12 11 10.0 & 14.29 & 0.13 & $+15.8 \ \pm$ 0.5 & 99 & 1 \\
K1095 & 21 30 20.32 & +12 00 42.4 & 12.67 & 0.64 & $-1.7 \ \pm$ 0.5 & 99 & 1,2,4 \\
K1096 & 21 30 20.45 & +12 17 55.9 & 14.03 & 0.60 & $-13.5 \ \pm$ 0.5 & 98 & 1,2 \\
\enddata
\tablenotetext{a}{2MASS coordinates \citep{skrutskie01}.}
\tablenotetext{b}{The visual photometry and membership probability from proper motions are taken from \citet{cudworth01}.}
\tablenotetext{c}{Average radial velocities were calculated from all cross-correlations.}
\tablenotetext{d}{Observations: 1: 2005 May 22, 2: 2006 May 11, 3: 2006 October 4, 4: 2006 October 7.}
\end{deluxetable*}

\clearpage

\begin{deluxetable*}{lcccc}
\tabletypesize{\scriptsize}
\tablecaption{H$\alpha$ Bisector Velocity of Cluster Members}
\tablewidth{0pt}
\tablehead{
\colhead{ID No.} &
\colhead{$v_{bis, 1}$ \tablenotemark{a}}	  & 
\colhead{$v_{bis, 2}$ \tablenotemark{a}}      & 
\colhead{$v_{bis, 3}$ \tablenotemark{a}} &
\colhead{$v_{bis, 4}$ \tablenotemark{a}} \\
\colhead{} & \colhead{\tiny{(km \ s$^{-1}$)}} & \colhead{\tiny{(km \ s$^{-1}$)}} & \colhead{\tiny{(km \ s$^{-1}$)}} &
\colhead{\tiny{(km \ s$^{-1}$)}}
}
\startdata
B5  & $-4.2 \ \pm$ 1.1 & \nodata & $-4.7 \ \pm$ 0.7   &  $-3.3 \ \pm$ 0.7   \\
B6  & $-1.8 \ \pm$ 0.7 & $-4.2        \ \pm$  0.7 & $-6.1 \ \pm$ 1.0   &  $-5.5 \ \pm$ 1.1  \\
B16 & \nodata & \nodata & $-2.7 \ \pm$ 0.7 & \nodata     \\
B30 & $-4.1 \ \pm$ 0.9 & $-2.5 \ \pm$		   1.3 & $-2.5 \ \pm$ 0.4    &  $-2.4 \ \pm$ 0.9    \\
C3  & $-1.8 \ \pm$ 1.0 & $-1.2 \ \pm$  0.7 &  $-0.9 \ \pm$ 0.5  &  $+0.3 \ \pm$ 0.7  \\
C20 & $+0.5 \ \pm$ 1.3 & $-0.4 \ \pm$  1.8 & $-0.5 \ \pm$ 1.0   &  $-0.2 \ \pm$ 2.0  \\
C35 & $-2.2 \ \pm$ 0.6 & \nodata & \nodata    &  \nodata   \\
GEB 254 & \nodata & \nodata & \nodata & $-8.5 \ \pm$ 0.9 \\
GEB 289 & \nodata & \nodata &  $-3.3 \ \pm$  0.6   & \nodata    \\
K12 & $-0.5 \ \pm$ 1.0 & $-1.7        \ \pm$  0.7 & $-0.6 \ \pm$ 0.5   & $-2.5 \ \pm$ 1.2   \\
K21 & \nodata & $-1.0  \ \pm$  0.8 & $-0.9 \ \pm$ 0.4   & $-1.7 \ \pm$ 0.5   \\
K22 & $+0.9 \ \pm$  1.1 & $-1.0 \ \pm$  1.0 &  $-1.8  \ \pm$ 0.3  & $+0.1 \ \pm$ 0.6   \\
K26 & $+0.1 \ \pm$ 0.8 & $+0.2  \ \pm$ 0.2 &  $+0.0  \ \pm$ 1.5  & $+0.1 \ \pm$ 1.2   \\
K27 & \nodata & $+1.2 \ \pm$  0.8 &  $-1.5 \ \pm$ 0.8  & $-1.2 \ \pm$ 0.5   \\
K31 & \nodata & $+0.9 \ \pm$  1.1 &  $-2.3 \ \pm$ 0.6  &  $-1.4 \ \pm$ 1.3   \\
K42 & \nodata & \nodata &  $-0.4 \ \pm$ 0.6    &   $+0.7 \ \pm$ 0.8    \\
K47 & $-1.5 \ \pm$ 0.7 & $-0.7        \ \pm$  0.4 & \nodata    &  \nodata   \\
K56 & $-2.2 \ \pm$ 0.9 & $+2.1 \ \pm$ 0.7 &  $+0.0 \ \pm$ 0.8  & $-1.7 \ \pm$ 0.9   \\
K60 & \nodata & \nodata & $+0.4 \ \pm$ 0.7    &  $+1.0 \ \pm$ 0.4     \\
K64 & $+0.5 \ \pm$ 1.0 & $+1.7 \ \pm$  1.3 & $-0.6 \ \pm$ 0.5   &  $+0.2 \ \pm$ 0.6  \\
K69 & $-0.6 \ \pm$ 0.5 & $-0.3        \ \pm$  0.3 & \nodata    &  $+0.4 \ \pm$ 0.7  \\
K70 & $-0.3 \ \pm$ 0.8 & \nodata & \nodata    &  \nodata   \\
K77 & $-1.8 \ \pm$ 0.6 & $-1.8        \ \pm$  0.6 & $-2.1 \ \pm$ 0.6   &  $-2.2 \ \pm$ 0.5  \\
K87 & \nodata & \nodata & \nodata & $-3.9 \ \pm$ 1.3 \\
K89 & \nodata & $-0.7  \ \pm$  0.7 & $-1.6 \ \pm$ 0.5   & $-0.8 \ \pm$ 0.5   \\
K92 & $-1.7 \ \pm$  1.2 & $+1.1 \ \pm$  0.5 &  $+0.5 \ \pm$ 0.7   &  $-1.0 \ \pm$ 1.2   \\
K105 & $-2.7 \ \pm$ 1.5 & \nodata & \nodata    &  $-2.0 \ \pm$ 1.2  \\
K112 & \nodata & \nodata & \nodata & $+0.9 \ \pm$ 1.6 \\
K114 & \nodata & \nodata &  $-1.7  \ \pm$ 0.4   &  \nodata     \\
K129 & \nodata & \nodata &  $+0.1 \ \pm$ 0.4    &  \nodata     \\
K133 & \nodata & $-0.8 \ \pm$  0.6 & $+0.5 \ \pm$ 1.6   &  \nodata   \\
K136 & \nodata & \nodata & \nodata & $-2.9 \ \pm$ 0.8 \\
K137 & \nodata & \nodata &  $+1.5 \ \pm$ 1.3      &  \nodata          \\
K144 & \nodata & $-5.3 \ \pm$  1.3 & \nodata    & $-5.7 \ \pm$ 0.9   \\
K145 & $+0.3 \ \pm$ 0.9 & $-0.2 \ \pm$  0.6 & $-0.4 \ \pm$ 1.2   & \nodata    \\
K146 & \nodata & $-2.5  \ \pm$  0.6 &  \nodata   &  \nodata   \\
K151 & \nodata & \nodata &  $+0.9 \ \pm$ 0.8     & \nodata      \\
K152 & $-0.4 \ \pm$ 1.0 & $-0.9 \ \pm$  0.8 &  $+0.2 \ \pm$ 1.1  & \nodata    \\
K153 & $+0.0 \ \pm$ 1.0 & \nodata & $+0.7 \ \pm$ 1.2   & $-0.4 \ \pm$ 1.5   \\
K158 & \nodata & $-11.1	    \ \pm$   1.0 & \nodata    &  \nodata   \\
K202 & \nodata & \nodata & $+0.1 \ \pm$ 0.8     &  \nodata    \\
K224 & $-2.9 \ \pm$ 0.4 & $-4.0       \ \pm$  0.6 &  $-4.5 \ \pm$ 0.9  & \nodata    \\
K238 & \nodata & \nodata & \nodata & $-3.1 \ \pm$ 0.6 \\
K255 & $-1.1\ \pm$  1.8 & $-1.0       \ \pm$  0.5 & \nodata    & \nodata    \\
K260 & $-3.0 \ \pm$ 1.2 & $-10.6	\ \pm$   0.9 & \nodata    &   \nodata  \\
K272 & \nodata & \nodata & $-3.0  \ \pm$ 1.6     &  \nodata     \\
K288 & $-0.91 \ \pm$ 0.82 & \nodata & $-2.7 \ \pm$ 0.6   & \nodata    \\
K328 & \nodata & $-2.4 \ \pm$ 0.6 & \nodata    &  $-3.3 \ \pm$ 1.3  \\
K337 & \nodata & $+0.1 \ \pm$ 0.6 & $-0.4  \ \pm$ 1.2   & $+2.9 \ \pm$ 1.6    \\
K341 & $-3.2 \ \pm$ 0.6 & $-6.9       \ \pm$  1.0 &  $-6.2 \ \pm$ 0.6  &   $-6.3 \ \pm$ 0.9 \\
K361 & \nodata & \nodata & $-1.9 \ \pm$ 0.4   &  \nodata    \\
K393 & $-2.0 \ \pm$ 0.9 & \nodata & \nodata    & \nodata    \\
K421 & \nodata & \nodata &  $-4.3 \ \pm$ 0.7     &  \nodata    \\
K431 & $-4.2 \ \pm$ 0.8 & $-5.0       \ \pm$  1.3 &  $-5.7 \ \pm$ 1.1  & \nodata    \\
K447 & \nodata & $-2.0 \ \pm$  0.6 &\nodata     &  \nodata   \\
K462 & \nodata & $-3.6 \ \pm$   0.7 & \nodata    & \nodata    \\
K476 & \nodata & $-0.2 \ \pm$  0.5 & \nodata    & \nodata    \\
K479 & \nodata & \nodata & \nodata & $-0.7 \ \pm$ 0.7 \\
K482 & \nodata & \nodata & $-6.2  \ \pm$ 1.2     & \nodata       \\
K506 & \nodata & \nodata & \nodata & $-0.6 \ \pm$ 1.9 \\
K550 & \nodata & $-1.3 \ \pm$  1.1 & $+1.6  \ \pm$ 0.4   & $+1.6 \ \pm$ 0.6   \\
K567 & \nodata & \nodata &  $-5.7  \ \pm$ 1.1   & \nodata      \\
K582 & $-13.2 \ \pm$ 1.6 & \nodata &  \nodata   & \nodata    \\
K583 & $-6.0 \ \pm$ 1.3 & \nodata &  \nodata   &  \nodata   \\
K647 & \nodata & \nodata & \nodata & $-6.2 \ \pm$ 0.8 \\
K654 & \nodata & \nodata & \nodata & $+0.1 \ \pm$ 0.7 \\
K672 & \nodata & \nodata &  $-4.6 \ \pm$ 0.9    &  \nodata     \\
K677 & \nodata & \nodata &  $-1.1 \ \pm$ 1.5   &  $-4.6  \ \pm$ 1.9    \\
K691 & $-1.2 \ \pm$  0.5 & $-0.6 \ \pm$        0.5 &  \nodata   &  \nodata   \\
K702 & $-6.1 \ \pm$ 0.8 & $-3.9       \ \pm$  0.9 & \nodata    & \nodata    \\
K709 & \nodata & \nodata & $-3.1 \ \pm$ 0.6     &  $-4.2  \ \pm$ 1.0     \\
K736 & \nodata & \nodata & $-2.1 \ \pm$ 0.3    &  \nodata    \\
K757 & $-2.8 \ \pm$ 0.5 & $-8.9       \ \pm$  1.1 & \nodata    & \nodata    \\
K800 & $+1.6 \ \pm$ 0.6 & \nodata & \nodata    &  \nodata   \\
K825 & \nodata & \nodata &  $+0.2 \ \pm$ 0.9      &  \nodata     \\
K846 & $-1.8 \ \pm$ 0.5 & \nodata &  \nodata   & $-3.3 \ \pm$ 1.2   \\
K853 & $-1.8 \ \pm$ 0.5 & $-2.2 \ \pm$ 0.8 & \nodata    & \nodata    \\
K866 & \nodata & \nodata &  $-1.3 \ \pm$ 0.4     & \nodata       \\
K875 & $-7.2 \ \pm$ 1.0 & $-4.4 \ \pm$ 0.4 &  \nodata   & \nodata    \\
K879 & $-1.3 \ \pm$ 0.5 & $-0.5 \ \pm$ 0.6 &  \nodata   &  \nodata   \\
K902 & \nodata & \nodata &  $-0.4 \ \pm$ 0.4    & $+0.8  \ \pm$0.5     \\
K906 & \nodata & \nodata &  $-0.9 \ \pm$ 0.4         &  \nodata       \\
K919 & $-2.4 \ \pm$ 0.5 & $-2.6 \ \pm$ 0.8 & \nodata    &  \nodata   \\
K925 & $-0.7 \ \pm$ 0.7 & $-0.4 \ \pm$ 0.8 &  \nodata   & \nodata    \\
K926 & \nodata & \nodata & \nodata & $+0.6 \ \pm$ 0.7 \\
K932 & \nodata & \nodata & \nodata & $-1.5 \ \pm$ 1.2 \\
K947 & \nodata & $-1.0 \ \pm$  1.1 & \nodata    &  \nodata   \\
K954 & \nodata & \nodata & \nodata & $-0.6 \ \pm$ 0.8 \\
K969 & $-4.0 \ \pm$ 0.6 & \nodata & \nodata    &  $-1.7 \ \pm$ 0.3  \\
K979 & $-6.7 \ \pm$ 1.3 & \nodata & \nodata    &  \nodata   \\
K989 & \nodata & \nodata &  $-0.4 \ \pm$ 1.0    & \nodata       \\
K993 & $-2.4 \ \pm$ 0.9 & \nodata &  \nodata   & \nodata    \\
K1010 & \nodata & \nodata & \nodata & $-2.3  \ \pm$ 1.6 \\
K1014 & \nodata & \nodata &  $+1.1 \ \pm$ 0.9   & \nodata     \\
K1029 & $-2.9 \ \pm$ 1.1 & $-3.0 \ \pm$ 1.0 & \nodata    & \nodata    \\
K1030 & $-1.3 \ \pm$ 0.5 & $-2.6 \ \pm$ 0.6 &  \nodata   & \nodata    \\
K1033 & $-0.9 \ \pm$ 0.5 & $+0.6  \ \pm$ 0.7 & \nodata    & \nodata    \\
K1040 & \nodata & \nodata &  $-6.3 \ \pm$ 1.4     &  \nodata      \\
K1049 & $-1.2 \ \pm$ 1.8 & \nodata &  \nodata   & \nodata    \\
K1054 & $-1.5 \ \pm$ 0.4 & $-1.1      \ \pm$  0.6 &  \nodata   & \nodata    \\
K1056 & \nodata & \nodata & \nodata & $-0.4 \ \pm$ 1.0 \\
K1069 & \nodata & \nodata & \nodata & $-1.5 \ \pm$ 0.8 \\
K1073 & $-3.6 \ \pm$ 0.8 & $-0.7      \ \pm$  0.7 & $+0.0 \ \pm$ 0.8   & $-1.9 \ \pm$ 0.9   \\
K1074 & $-0.1 \ \pm$ 0.9 & \nodata & $-0.5 \ \pm$ 1.6   &  $-0.6  \ \pm$ 1.1  \\
K1079 & $-1.4 \ \pm$  0.4 & $-0.9 \ \pm$       0.7 & \nodata    & \nodata    \\
K1083 & $-0.8 \ \pm$ 1.6 & $-1.6      \ \pm$  1.6 & $+1.5 \ \pm$ 1.2   &  \nodata   \\
K1084 & $-0.7 \ \pm$ 0.8 & $-0.6 \ \pm$        0.6 & $-1.1 \ \pm$ 0.5   & \nodata    \\
K1097 & $+2.5 \ \pm$ 0.8 & $+0.6 \ \pm$  0.8 & $-1.0 \ \pm$ 0.7    &  $-1.1 \ \pm$ 0.6  \\
K1106 & \nodata & $-0.2	    \ \pm$  0.7 & $-1.7 \ \pm$ 1.5   & \nodata    \\
K1136 & $-1.0 \ \pm$ 1.4 & $-0.2      \ \pm$ 0.7 &  $+0.9 \ \pm$ 0.7  &  $-1.3 \ \pm$ 1.0  \\
\enddata
\tablenotetext{a}{Observations: 1: 2005 May 22; 2: 2006 May 11; 3: 2006 October 4; 4: 2006 October 7.}
\end{deluxetable*}

\clearpage

\begin{deluxetable*}{lc}
\tabletypesize{\scriptsize}
\tablecaption{Parameters of \ion{Ca}{2}~K Line for Stars Showing Emission}
\tablewidth{0pt}
\tablehead{
\colhead{ID No.}           & \colhead{$B/R$ 2005 May 23}
}
\startdata
K77 & $>1$ \\
K224 & $<1$ \\
K260 & $>1$ \\
K341 & $>1$ \\
K393 & $>1$ \\
K582 & $<1$ \\
K702 & not clear \\
K757 & $<1$ \\
K853 & $>1$ \\
K875 & $~1$ \\
K879 & $>1$ \\
K969 & $~1$ \\
K979 & not clear \\
K1029 & not clear \\
\enddata
\tablecomments{The parameter $B/R$ is the intensity 
ratio of Blue (short wavelength) and Red (long wavelength) emission peaks.}
\end{deluxetable*}

\end{document}